\documentclass[12pt]{article}
\pdfoutput=1
\usepackage{jheppub}
\allowdisplaybreaks

\newcommand{\cn}{{\cal N}}
\newcommand{\cl}{{\cal L}}
\newcommand{\cf}{{\cal F}}
\newcommand{\cj}{{\cal J}}
\def\bal#1\eal{\begin{align}#1\end{align}}
\def\>{\rangle}
\def\<{\langle}
\def\b{\beta}
\def\c{\chi}
\def\e{\epsilon}

\def\w{\omega}
\def\p{\pi}
\def\q{\theta}
\def\x{\xi}
\def\S{\Sigma}
\def\pa{\partial}

\title{A Supersymmetric SYK-like Tensor Model}

\author{Cheng Peng, Marcus Spradlin and Anastasia Volovich}
\affiliation{Department of Physics, Brown University, Providence RI 02912, USA}
\emailAdd{cheng$\underline{~}$peng@brown.edu, marcus$\underline{~}$spradlin@brown.edu, anastasia$\underline{~}$volovich@brown.edu}

\abstract{
We consider a supersymmetric SYK-like model without quenched disorder that is built by coupling
two kinds of
fermionic $\cn=1$
tensor-valued superfields, ``quarks'' and ``mesons''.
We prove that the model has a well-defined large-$N$ limit in which the (s)quark 2-point functions
are dominated by mesonic ``melon'' diagrams.
We sum these diagrams to obtain the Schwinger-Dyson equations and show that
in the IR, the solution agrees with that of the supersymmetric SYK model.
}

\begin{document}

\maketitle

\section{Introduction}

The SYK model~\cite{Sachdev:1992fk,PG,GPS,KitaevTalk1,KitaevTalk2}
has elicited considerable attention in the recent high energy literature
for being an apparently solvable
model~\cite{Polchinski:2016xgd,Garcia-Garcia:2016mno,Jevicki:2016ito,Maldacena:2016hyu,Jevicki:2016bwu}
that encapsulates non-trivial
features of black
holes~\cite{Danshita:2016xbo,Garcia-Alvarez:2016wem,Shenker:2013pqa,Shenker:2014cwa,Sachdev:2015efa,Maldacena:2015waa}
with AdS${}_2$
horizons~\cite{Strominger:1998yg,Maldacena:1998uz,Sachdev:2010um,Almheiri:2014cka,Engelsoy:2016xyb, Jensen:2016pah, Maldacena:2016upp,Cvetic:2016eiv,Grumiller:2016dbn,DFGGJS}.
Several extensions of the model with various interesting properties also have been proposed~\cite{Gu:2016oyy,Berkooz:2016cvq,Gross:2016kjj,Cotler:2016fpe}.
In particular, a supersymmetric generalization has recently been studied by Fu, Gaiotto, Maldacena, and Sachdev (FGMS)~\cite{Fu:2016vas}  (see also similar studies of supersymmetric lattice models in e.g.~\cite{Nicolai:1976xp,Fendley:2002sg,Huijse:2008rwp,Anninos:2016szt,Sannomiya:2016mnj}).
Some recent studies of the SYK model in the CMT literature
include~\cite{Bagrets:2016cdf,Fu:2016yrv,You:2016ldz,Banerjee:2016ncu}.

In~\cite{Witten:2016iux} Witten constructed an SYK-like model that
does not involve averaging over a random coupling.
The model is based on a certain tensor model due to Gurau and
collaborators, on
which there
exists an extensive literature (see e.g.~\cite{Gurau:2012ix,Gurau:2010ba,Gurau:2016cjo,Gurau:2009tw,Gurau:2011xp,Gurau:2011aq,Gurau:2011xq,Bonzom:2011zz,Bonzom:2012hw} and references therein).
Witten showed that the large-$N$ limit of this model has the same correlation functions and thermodynamics as the SYK model.
The $1/N$ expansion of the Gurau-Witten model has been explicitly constructed in~\cite{Gurau:2016lzk}, and a generalization that shares many of
its salient features but is based on an ``uncolored'' tensor model
has been proposed and studied in great detail by
Klebanov and Tarnopolsky~\cite{Klebanov:2016xxf},
building on an earlier analysis of this type of model
in~\cite{Carrozza:2015adg}.

The absence of quenched disorder is an attractive
feature which means that the model of~\cite{Witten:2016iux} is a true quantum
theory instead of an average over an ensemble of theories, as in the original
SYK model.
It also provides possibilities for future studies of other chaotic systems
that manifestly avoid spin-glass phases,
since the replica symmetry is not present.
(An alternative SYK-like model without disorder has been proposed in~\cite{NT}.)
Given these advantages, it is natural to ask if we can find tensor models that describe other quenched disordered systems with properties similar to the SYK model.
One example of such a system is the supersymmetric SYK model
of FGMS~\cite{Fu:2016vas}.
This model was shown to be chaotic at late times and has two towers of
higher spin operators, together with their superpartners, in its spectrum.

Motivated by both~\cite{Fu:2016vas} and~\cite{Witten:2016iux}, in this paper
we propose a tensor model
that mirrors the supersymmetric FGMS model.
We promote the
fermionic ``quark'' fields of~\cite{Witten:2016iux}
to $\cn=1$ fermionic tensor-valued superfields.
Like the model of~\cite{Witten:2016iux} our model has
global $O(n)^6/\mathbb{Z}_2^2$ symmetry.  We introduce additional
``meson'' superfields in order to be able to construct
a supercharge that is cubic in the quark and meson fields.
The Hamiltonian is the square of the supercharge.
We prove that this model has a well-defined large-$n$ limit.
Quark loops are suppressed by $1/n$, and the dominant
mesonic exchange graphs involve ``melon'' diagrams of a type
that are very familiar in tensor models.
The nontrivial effective dynamics of the quarks is completely
due to having to propagate through the
``mesonic melon patch.''
We sum the relevant melonic diagrams in the large-$n$ limit and
show that the solution to the Schwinger-Dyson equations in the
IR limit agrees with that of the supersymmetric FGMS model.
In particular, the dimension of the quark field is $\Delta = 1/6$
as in~\cite{Fu:2016vas}. As a result, we conclude that other features
such as the chaotic behavior and the operator spectrum in the
OPEs are identical to those in~\cite{Fu:2016vas}.

It is certainly also interesting to
ask whether there exists a supersymmetrization
of the Gurau-Witten
model that has the same IR physics as the fermionic SYK model
(in which, in particular, the fermion field has IR dimension $\Delta = 1/4$).
In our model, we find that taking the large-$n$ limit does not commute
with integrating out the auxiliary bosons. If we first perform the
latter, then with a judicious choice of coupling constant the meson
fields decouple completely. This process does recover the interaction
\begin{equation}
j\, \psi_0^{i_{01}i_{02}i_{03}}\psi_1^{i_{01}i_{12}i_{13}}\psi_2^{i_{02}i_{12}i_{23}}\psi_3^{i_{03}i_{13}i_{23}}\label{hw}
\end{equation}
that was shown to be diagrammatically equivalent to SYK
in the large-$n$ limit when $j \sim n^{-3/2}$~\cite{Witten:2016iux}.
However, it also generates simultaneously the only other quartic operator in the colored
SYK-like tensor model,
\begin{equation}
g\, (\psi_a\psi_b)(\psi_a\psi_b) = g\,( \psi_0^{i_{01} i_{02} i_{03}}
\psi_1^{i_{01} i_{12} i_{23}} \psi_0^{j_{01} i_{02} i_{03}}
\psi_1^{j_{01} i_{12} i_{23}} + \cdots)\,,\label{hw2}
\end{equation}
where the $\cdots$ represents a sum over $a, b \in \{0,1,2,3\}$.
This kind of operator is referred to as a ``pillow" operator in the recent study of the uncolored SYK-like tensor model by Klebanov and Tarnopolsky~\cite{Klebanov:2016xxf}. The colored version of this pillow operator is also mentioned in~\cite{Klebanov:2016xxf}. An earlier thorough study of models with pillow operators can be found in~\cite{Carrozza:2015adg}.
However, diagrams involving this pillow operator scale differently than those without it, suggesting a different large-$n$ limit.

Therefore our model is better regarded as a tensor version of the
supersymmetric FGMS model, rather than an honest supersymmetrization
of the SYK-like Gurau-Witten tensor model.
Certainly there could be more than one way to supersymmetrize the fermionic tensor models, and there could be more than one tensor model that resembles the SYK-like models, as demonstrated for example by the uncolored model studied in~\cite{Klebanov:2016xxf}.

In section~1 we review the fermionic SYK-like Gurau-Witten tensor model of~\cite{Bonzom:2011zz,Witten:2016iux},
the supersymmetric SYK model of FGMS~\cite{Fu:2016vas}, and
discuss their large-$N$ limits. In sections~2 through~4
we present and discuss our proposal for the supersymmetric tensor model and discuss its large-$N$
limit, with most of the technical
details tucked away safely in appendix~A.

\subsection{Review of the SYK-like tensor model (Gurau-Witten)}
\label{wittenreview}

The SYK-like tensor model discussed in~\cite{Witten:2016iux} is a modification of the construction of~\cite{Bonzom:2011zz} and is based on $q = D + 1$ real tensor-valued fermions
\begin{equation}
\psi_a^{a_1\ldots a_D}\,,
\qquad a = 0, 1, \ldots, D
\label{eq:psidef}
\end{equation}
in 0$+$1 dimensions.
Here each $a_i$ is a vector index of a distinct $O(n)$ group and we focus on the case $D=3$ that is relevant to the SYK model. In this case each fermion is charged under 3 different $O(n)$ groups and hence has 3 vector indices. Specifically these take the form
\begin{equation}
\psi_0^{i_{01}i_{02}i_{03}}, \qquad
\psi_1^{i_{01}i_{12}i_{13}}, \qquad
\psi_2^{i_{02}i_{12}i_{23}}, \qquad
\psi_3^{i_{03}i_{13}i_{23}}\,,
\end{equation}
where each $i_{ab}$ is a vector index of an $O(n)$ group $G_{ab}$. There are a total of $\frac{D(D+1)}{2}=6$ distinct $O(n)$ groups. The interaction Hamiltonian
has a global $O(n)^6/\mathbb{Z}_2^2$ symmetry:
\begin{equation}
H^{\text{int}}= j\psi_0^{i_{01}i_{02}i_{03}}\psi_1^{i_{01}i_{12}i_{13}}\psi_2^{i_{02}i_{12}i_{23}}\psi_3^{i_{03}i_{13}i_{23}}\,.
\label{eq:hamiltonian}
\end{equation}
Here $j$ is a coupling constant and the repeated $O(n)$ indices are summed over. Altogether there are
\begin{equation}
N=4n^3
\end{equation}
fermionic degrees of freedom.

\begin{figure}[t]
\centering
\includegraphics[width=1.0\linewidth]{
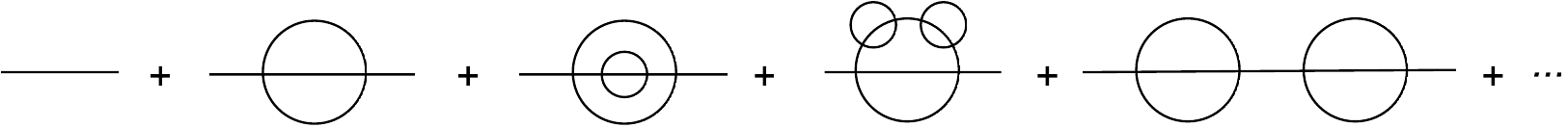}
\caption{The large-$N$ limit of the SYK-like model of~\cite{Witten:2016iux}, like many tensor models, is dominated by ``melon'' graphs. Here we show several melonic contributions to the 2-point function. In general they may be generated iteratively using the basic building block shown in the second term.}
\label{fig:melon}
\end{figure}

Products of fermions can be represented as 3-valent graphs with four types of labeled vertices $V_0, \ldots, V_3$. Specifically,
\begin{itemize}
\item each $\psi_a$ is represented by a 3-valent vertex of type $V_a$;
\item an unoriented edge $\overline{ab}$ connecting two vertices of types $V_a$ and $V_b$ corresponds to a contraction over the $O(n)$ index $i_{ab}$;
\item and scalars under the full symmetry group correspond to graphs with no open edges.
\end{itemize}
It is well-known (see for example~\cite{Gurau:2010ba,Gurau:2011aq,Gurau:2011xq,Bonzom:2011zz,Bonzom:2012hw,Gurau:2016cjo}) that the tensor models on which~\cite{Witten:2016iux} is based are dominated in the large-$N$ limit by a certain class of ``melon'' diagrams (see figure~\ref{fig:melon}). The quantum mechanical tensor model~\eqref{eq:hamiltonian} has a well-defined large-$N$ limit provided that the coupling $j$ is taken to scale as
\begin{equation}
j \sim n^{-3/2} \sim N^{-1/2}\,.
\end{equation}
In~\cite{Witten:2016iux} it was shown that the melon diagrams which dominate the model in this limit are the same as the Feynman diagrams that dominate the large-$N$ limit of the SYK model. The Schwinger-Dyson equations of the two theories are therefore formally identical (except for extra copies in the tensorial model due to the vector indices). As a result, the correlation functions, chaotic behavior, etc.~of the two models should be identical.

\subsection{Review of the supersymmetric SYK model (FGMS)}
\label{susyrev}

A direct supersymmetrization of the SYK model was constructed
by FGMS~\cite{Fu:2016vas} using superfields.
In $0+1$ dimensions, an $\cn=1$ superfield can be constructed with the help of a single real Grassmann variable $\q$. For instance, a fermionic superfield is defined as
\begin{equation}
\Psi(t,\q)=\psi(t)+\q \, b(t)\,,
\end{equation}
where $\psi(t)$ is fermionic and $b(t)$ is bosonic. The supersymmetry transformation is generated by the off-shell supercharge
\begin{equation}
Q=\pa_\q- { i} \q \, \pa_t\,,
\end{equation}
which satisfies $Q^2=- { i}\pa_t$. We further define the super-derivative
\begin{equation}
D=\pa_\q + { i}\q \, \pa_t\,, \qquad D^2={ i}\pa_t\,,
\end{equation}
which anticommutes with the supercharge
\begin{equation}
\{Q,D\}=0\,.
\end{equation}
The supersymmetry transformation of the superfield is
\begin{equation}
\delta_\x \psi(t)+\q \, \delta_\x b(t) \equiv \delta_\x \Psi(t,\q)=\x Q \Psi(t,\q)= \x b(t) -{ i}\x \q\, \pa_t \psi(t)\,,
\end{equation}
from which we read off the transformations of the components
\begin{equation}
\delta_\x \psi(t)=\x b(t)\,, \qquad \delta_\x b(t) = { i}\x \pa_t \psi(t)\,,
\end{equation}
which in turn imply
\begin{equation}
Q\psi(t)=b(t)\,, \qquad Q b(t) = { i}\pa_t \psi(t)\,.
\label{Qvar}
\end{equation}
In $0+1$ dimensions any $\cn=1$ superfield can be decomposed into a constant piece and a piece proportional to $\q$. From the form of the supercharge, the $\q$-dependent piece always transforms into a total derivative. As a result, any integral of the form
\begin{equation}
\int d\q\, dt\, f(t,\q)
\end{equation}
is manifestly invariant under supersymmetry since the $d\theta$ integral picks out the term proportional to $\q$ that transforms into a total derivative. The simplest interacting supersymmetric Lagrangian for a collection of superfields $\Psi_i$ takes the form
\begin{equation}
\cl = \int d\q \left(-\frac{1}{2}\Psi_i D \Psi_i + { i} \frac{C_{ijk}}{3} \Psi_i\Psi_j\Psi_k\right),
\label{superspaceL}
\end{equation}
where repeated indices $i, j, k$ are summed over and $C_{ijk}$ is a totally antisymmetric coupling constant. In component form, this gives
\begin{equation}
\cl = \frac{{i}}{2}\psi_i \pa_t \psi_i-\frac{1}{2} b_ib_i + {i} C_{ijk} b_i\psi_j\psi_k\,.
\end{equation}
Notice that there is no kinetic term for the $b_i$ field, so it is auxiliary and can be integrated out. Substituting the resulting constraint
\begin{equation}
b_i=i C_{ijk} \psi_j\psi_k
\label{bpsi}
\end{equation}
back into the Lagrangian gives
\begin{equation}
\cl = \frac{{i}}{2}\psi_i \pa_t \psi_i-\frac{C_{ijk}C_{imn}}{2} \psi_j\psi_k \psi_m\psi_n\,.
\label{eq:integratedout}
\end{equation}
As noted in~\cite{Fu:2016vas} this is very similar to the original SYK model except that the fundamental coupling is not a direct 4-point coupling $J_{ijkl}$ but a Yukawa type $C_{ijk}$ leading to an effective 4-point coupling $\sim C_{ijm}C_{klm}$. In the SYK model each $J_{jkmn}$ is a Gaussian random variable, but if instead the Yukawa couplings $C_{ijk}$ are drawn from a Gaussian distribution then it changes the structure of the large-$N$ equations. As a result, the scaling dimension of the fields in the IR of the supersymmetric model are different~\cite{Fu:2016vas} than those of the original fermionic SYK model.

\subsection{Large-$N$ limit of the supersymmetric model}\label{lnsec}

Most of the analysis in~\cite{Maldacena:2016hyu,Fu:2016vas} was done using effective actions obtained by averaging over the random couplings.
This approach is convenient because the Schwinger-Dyson equations of the original theory become the classical equations of motion of the effective/collective fields. Correlation functions can then be computed using the solutions of these equations of motion.

\begin{figure}[t]
\centering
\includegraphics[width=0.5\linewidth]{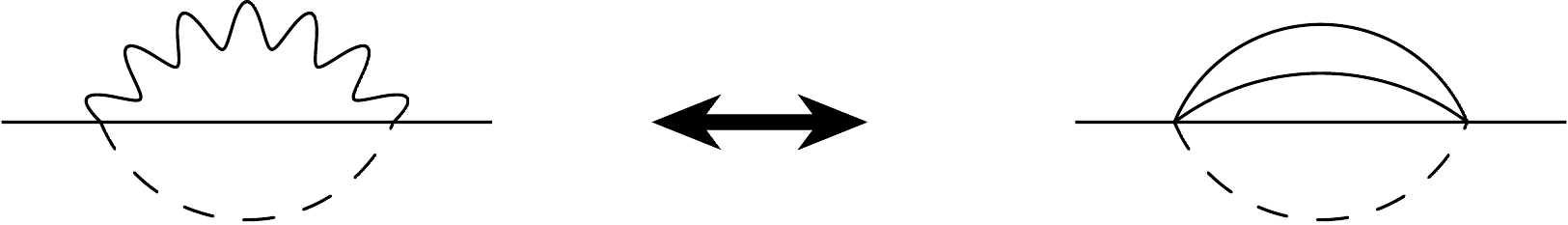}
\caption{Left: the dominant ``core" diagram contributing to the $\psi\psi$ 2-point function in the supersymmetric FGMS model. Right: the corresponding dominant ``core" diagram in the fermionic SYK model. In both graphs the solid lines represent $\psi_i$ fields, the wavy lines represent $b_i$ fields, and the dashed lines represent a correlation of a product of Gaussian random couplings; $C_{ijk}$ on the left and $J_{ijkl}$ on the right. The full set of dominant graphs is generated iteratively using this core.}
\label{fig:susy2pt}
\end{figure}

The effective action and Schwinger-Dyson equations of the supersymmetric model are written down in eqs.~(2.10) and~(2.11) of~\cite{Fu:2016vas}. Since the Schwinger-Dyson equations can be visualized diagrammatically, e.g. as in figures~1 and~2 of~\cite{Maldacena:2016hyu}, we can translate the first line of equation (2.11) in~\cite{Fu:2016vas} back to a diagrammatic presentation which simply states that the leading large-$N$ contributions to the $\langle \psi_i(t_1)\psi_i(t_2)\rangle$ 2-point function come from all ``melon" diagrams that can be constructed from the building block shown in the left panel of figure~\ref{fig:susy2pt}. These diagrams can also be shown to be dominant by a straightforward power counting argument. If we use the constraint~(\ref{bpsi}) to replace the bosonic field by a pair of fermionic fields, together with a change of the vertices to those of the Lagrangian~(\ref{eq:integratedout}) with the $b_i$ integrated out, this diagram precisely reproduces the dominant ``core" diagram of the $\langle \psi_i(t_1)\psi_i(t_2)\rangle$ 2-point function in the fermionic SYK model, shown in the right panel of figure~\ref{fig:susy2pt}.

\begin{figure}[t]
\centering
\includegraphics[width=0.5\linewidth]{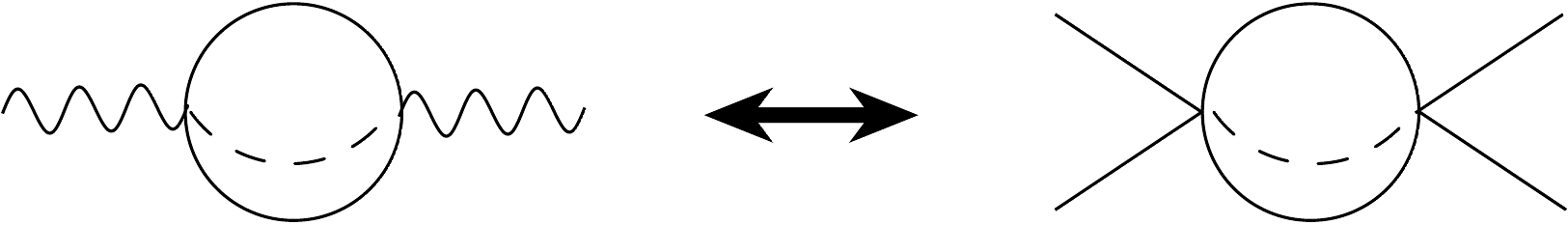}
\caption{Left: the dominant ``core" diagram contributing to the $bb$ 2-point function in the supersymmetric FGMS model. Right: the dominant ladder ``rung" diagram contributing to the $\psi\psi\psi\psi$ 4-point function in the fermionic SYK model. The full set of dominant graphs is generated iteratively using this core (with propagator corrections obtained by also iterating the right panel of figure~\ref{fig:susy2pt}).}
\label{fig:susy4pt}
\end{figure}

Similarly, the second line of eq.~(2.11) in~\cite{Fu:2016vas} implies that the leading contributions to the $\langle b_i(t_1) b_i(t_2)\rangle$ 2-point function come from ``melon" diagrams of the building block drawn in the left panel of figure~\ref{fig:susy4pt}. If we use the constraint~(\ref{bpsi}) and change the vertices accordingly again, this diagram precisely reduces to the dominant ``kernel" diagram of the 4-point functions in the fermionic SYK model, as shown for example in figure~3 of~\cite{Maldacena:2016hyu} and in the right panel of our figure~\ref{fig:susy4pt}.

Note that this analysis does not mean that a 2-point function of the $b_i$ fields in the supersymmetric model should be thought of as a 4-point function of $\psi_i$'s in the fermionic model. Rather, this merely demonstrates that the dominant diagrams in the supersymmetric model are pictorially the same set of dominant diagrams as in the fermionic model. Specifically, we can resolve bosonic lines into a pair of fermionic lines to see how the dominant diagrams map from the supersymmetric model into the fermionic model. We will observe a similar phenomenon in our proposed supersymmetric tensor model.

\section{A supersymmetric tensor model}
\label{stmodel}

We now combine the ideas reviewed in sections~\ref{wittenreview} and~\ref{susyrev} to construct a tensor model version of the supersymmetric FGMS model. To this end we start by promoting the tensor-valued fermion $\psi_a$ of eq.~(\ref{eq:psidef}) to the tensorial superfield
\begin{equation}
\Psi_a^{\underline{i_a}}=\psi_a^{\underline{i_a}}+\q \,\b_a^{\underline{i_a}}\,,
\label{eqn:tsf}
\end{equation}
where $\underline{i_a}$ is a collective notation for the vector indices of the individual $O(n)$ groups. A naive supersymmetrization of the tensorial model would be to replace the $\Psi(t,\q)$ superfields in the action~\eqref{superspaceL} by the tensorial version~\eqref{eqn:tsf}. However, one immediately encounters the problem that it is not possible to construct a scalar under all of the $O(n)$ groups that is cubic in the tensorial fields $\Psi_a^{\underline{i_a}}$ of the $q=4$ model.

We can compensate for this problem by introducing a set of ``meson'' superfields
\begin{equation}
\Pi_{ab}^{i_{a0}\ldots \widehat{i_{aa}} \ldots \widehat{i_{ab}} \ldots i_{a3},\,i_{b0}\ldots \widehat{i_{bb}} \ldots \widehat{i_{ab}} \ldots i_{b3}}\equiv \Pi_{ab}^{\underline{i_a},\,\underline{i_b}}=\c_{ab}^{\underline{i_a},\,\underline{i_b}}+\q \,\pi_{ab}^{\underline{i_a},\,\underline{i_b}}\label{mesonsuperfield}
\end{equation}
with fermionic components $\chi_{ab}^{\underline{i_a},\,\underline{i_b}}$ and bosonic components $\pi_{ab}^{\underline{i_a},\,\underline{i_b}}$. The notation $\widehat{i_{ab}}$ means that the vector index associated to the group $G_{ab} $ is missing. All fields are antisymmetric in $a \leftrightarrow b$ so, to be completely explicit, the $\frac{q(q-1)}{2}=6$ mesonic superfields have the form
\begin{align}
\Pi_{01}^{i_{02}i_{03},\,i_{12}i_{13}}, \qquad
\Pi_{02}^{i_{01}i_{03},\,i_{12}i_{23}}, \qquad
\Pi_{03}^{i_{01}i_{02},\,i_{13}i_{23}},\\
\Pi_{12}^{i_{01}i_{13},\,i_{02}i_{23}}, \qquad
\Pi_{13}^{i_{01}i_{12},\,i_{03}i_{23}}, \qquad
\Pi_{23}^{i_{02}i_{12},\,i_{03}i_{13}}\,.
\end{align}

Now we can write down an $O(n)^6/\mathbb{Z}_2^2$-invariant interacting Lagrangian by appropriately coupling the $\Psi_a^{\underline{i_a}}$ ``quark'' fields to the $\Pi_{ab}^{\underline{i_a},\,\underline{i_b}}$ meson fields:
\begin{equation}
\cl= \int d\q \left(-\frac{1}{2}\Psi_a D \Psi_a
-\frac{1}{2}\Pi_{ab} D {\Pi}_{ab}
+i \frac{g}{2} {\Pi}_{ab}(\Psi_a\Psi_b+\frac{1}{2}\e_{abcd} \Psi_c\Psi_d)+ i \frac{h}{6} \Pi_{ab}\Pi_{bc}
{\Pi}_{ca}\right)\,,
\label{supertensorLg}
\end{equation}
where $g$, $h$ are coupling constants and the indices $a, b, c$ are summed over. In terms of component fields, the above Lagrangian reads
\begin{multline}
 \cl = \frac{i}{2}\psi_a\pa_t \psi_a-\frac{1}{2}\b_a\b_a  + \frac{i}{2}\c_{ab} \pa_t {\c}_{ab}-\frac{1}{2}\p_{ab}{\p}_{ab}
\\ + i \frac{g}{2} {\p}_{ab}(\psi_a\psi_b +\frac{1}{2}\e_{abcd} \psi_c\psi_d) - i g {\c}_{ab} (\b_a \psi_b+\frac{1}{2}\e_{abcd} \b_c \psi_d) + i \frac{h}{2} {\p}_{ab}\c_{bc}\c_{ca} \ .
\end{multline}

Next let us compare this supersymmetric tensor model to the fermionic tensor model of~\cite{Witten:2016iux}. To achieve this we integrate out the non-dynamical bosonic fields $\b_a$ and $\p_{ab}$.
They are determined to satisfy the constraints
\bal
\b_a&= -i g \big({\c}_{ab}\psi_b+\frac{1}{2}\e_{abcd} \c_{bc}\psi_d\big)\,,\\
{\p}_{ab}&=i \frac{g}{2} \big(\psi_a\psi_b+\frac{1}{2}\e_{abcd}\psi_c\psi_d\big)+ i \frac{h}{2} \c_{bc} \c_{ca}   \label{eq:constraint}\,.
\eal
After substituting these values back into the Lagrangian we arrive at
\begin{multline}
 \cl = \frac{i}{2}\psi_a\pa_t \psi_a+ \frac{i}{2}\c_{ab} \pa_t {\c}_{ab} +\left(\frac{g h}{8}-\frac{g^2}{2}\right)\big( 2{\c}_{ab}{\c}_{ac}\psi_c\psi_b+\e_{abcd}\c_{bc}\c_{ae}\psi_e\psi_d\big) \\- \frac{g^2}{8} \e_{abcd}\psi_a\psi_b\psi_c\psi_d - \frac{g^2}{4} (\psi_a\psi_b)(\psi_a\psi_b) +  \frac{h^2}{8}  \c_{ab} \c_{bc}\c_{cd}\c_{da}\,.
\end{multline}
Evidently the Lagrangian simplifies if we choose
\begin{equation}
h=4g\,.
\label{hg}
\end{equation}
In the next section we discuss the significant ramifications of this choice for the large-$N$ limit; not because of the precise factor of $4$ but because eq.~(\ref{hg}) ties together the scaling of $h$ and $g$ with $n$. We expect the large-$N$ limit of this theory to be qualitatively insensitive to the precise numerical value of $h/g$, but the choice $4$ is clearly appealing because it leads to completely decoupled Lagrangians
\begin{align}
\label{Lsimple}
\cl_\psi &= \frac{i}{2}\psi_a\pa_t \psi_a- \frac{g^2}{8} \e_{abcd}\psi_a\psi_b\psi_c\psi_d - \frac{g^2}{4} (\psi_a\psi_b)(\psi_a\psi_b)\,,\\
\cl_\chi &= \frac{i}{2}\c_{ab} \pa_t {\c}_{ab}
+ 2{g^2} \c_{ab} \c_{bc}\c_{cd}\c_{da}
\label{eq2}
\end{align}
for the quarks and mesons.

The second interaction term
\begin{equation}
\label{eq:newinteraction}
(\psi_a\psi_b)(\psi_a\psi_b) = \psi_0^{i_{01} i_{02} i_{03}}
\psi_1^{i_{01} i_{12} i_{23}} \psi_0^{j_{01} i_{02} i_{03}}
\psi_1^{j_{01} i_{12} i_{23}} + \cdots
\end{equation}
in eq.~(\ref{Lsimple}) has different index structure as the first interaction term in that equation
\begin{equation}
\label{eq:witteninteraction}
\psi_0^{i_{01}i_{02}i_{03}}\psi_1^{i_{01}i_{12}i_{13}}\psi_2^{i_{02}i_{12}i_{23}}\psi_3^{i_{03}i_{13}i_{23}}\,,
\end{equation}
which is precisely the one considered in~\cite{Witten:2016iux}.

Let us record here for later use the Lagrangian at the value~(\ref{hg}) both in terms of superfields
\begin{equation}
 \cl= \int d\q \left(-\frac{1}{2}\Psi_a D \Psi_a
-\frac{1}{2}\Pi_{ab} D {\Pi}_{ab}
+i \frac{g}{2} {\Pi}_{ab}(\Psi_a\Psi_b+\frac{1}{2}\e_{abcd} \Psi_c\Psi_d)+ i \frac{2g}{3} \Pi_{ab}\Pi_{bc}
{\Pi}_{ca}\right)\,,
\label{supertensorLhg}
\end{equation}
and in terms of the component fields,
\begin{multline}
\cl = \frac{i}{2}\psi_a\pa_t \psi_a-\frac{1}{2}\b_a\b_a  + \frac{i}{2}\c_{ab} \pa_t {\c}_{ab}-\frac{1}{2}\p_{ab}{\p}_{ab}
\\ + i \frac{g}{2} {\p}_{ab}(\psi_a\psi_b +\frac{1}{2}\e_{abcd} \psi_c\psi_d) - i g {\c}_{ab} (\b_a \psi_b+\frac{1}{2}\e_{abcd} \b_c \psi_d) +  2 i g {\p}_{ab}\c_{bc}\c_{ca} \ .\label{Lcmp}
\end{multline}

\section{Large-$N$ limit of the supersymmetric tensor model}

In this section we analyze the large-$N$ limit of the Lagrangian~(\ref{Lcmp}) directly, without first integrating out the bosonic fields. Instead of using the effective action approach of~\cite{Maldacena:2016hyu,Fu:2016vas} that is very natural for taking the disorder average, we directly analyze the dominant Feynman diagrams in the large-$N$ limit and study the low energy behavior by deriving and solving the appropriate Schwinger-Dyson equations.

In our model the total number of degrees of freedom is
\begin{equation}
N=4n^3+6n^4 = \mathcal{O}(n^4)\,,\label{Nn}
\end{equation}
which is obviously larger than the $\mathcal{O}(n^3)$ of the purely fermionic model. In the following, we will use the word ``large-$n$" simply because we will be counting directly powers of $n$, not $N$. The main technical result of our paper is that a sensible large-$n$ limit of the supersymmetric tensor model~(\ref{Lcmp}) exists if we scale the coupling constant as
\begin{equation}
g \sim n^{-1}\,.\label{gn}
\end{equation}
The detailed proof of this is relegated to the appendix; we now proceed with several comments about its implications.

First we note that with $g \sim 1/n$, the large-$n$ expansion will manifestly have an expansion in inverse integer powers of $n$, being the rank of each of the six global $O(n)$ groups, similar to how the expansion is organized in the approach of~\cite{Klebanov:2016xxf}. On the other hand, we could translate our large-$n$ expansion into a large-$N$ expansion using eq.~(\ref{Nn}), which would lead to fractional powers of $1/N$ as noted in~\cite{Witten:2016iux} for the fermionic tensor model.

Next let us note that $g \sim 1/n$ implies a different scaling $g^2 \sim 1/n^2$ for the coefficient of the quartic interaction~(\ref{eq:newinteraction}) compared to the scaling $1/n^{3/2}$ which was found in~\cite{Witten:2016iux} to give a sensible and non-trivial scaling limit of the purely fermionic model with interaction~(\ref{eq:witteninteraction}). This change in the necessary scaling behavior is qualitatively similar to what happens in the supersymmetric FGMS model, see e.g.~the discussion below eq.~(1.5) of~\cite{Fu:2016vas}. However, the difference seems to be more significant in our case; unlike the case in~\cite{Fu:2016vas}, the change in the large-$n$ scaling here is inherently due to the introduction of the meson fields which dominate the large-$n$ limit. This fact is of course evident from eq.~(\ref{Nn}) but it will also be seen more explicitly in the appendix. Indeed, in the strict large-$n$ limit, correlation functions of $\psi_a$ fields are dominated by graphs with no $\psi_a$ loops (see for example the top panel of figure~\ref{fig:sd1}). The nontrivial effective dynamics of the quark field $\psi_a$ is therefore completely a consequence of it having to propagate through a background of $\pi_{ab}$ fields like some kind of ``mesonic melon patch''.

Let us also note that we can define the model~\eqref{Lcmp} in more than zero spatial dimensions, and all of the large-$n$ diagrammatic analysis in the following and in the appendix applies identically in these higher dimension models.

Finally, for the purpose of large-$n$ power counting, we do not distinguish between the $\psi_a$ and $\b$ fields; they give identical factors of $n$. For the same reason we make no distinction between the $\chi_{ab}$ and $\pi_{ab}$ fields. In figures such as~\ref{fig:sd1} we will use single lines to denote the $\psi_a, \b_a$ fields and double lines to denote the $\chi_{ab}, \pi_{ab}$ fields. Essentially, this means that we draw superspace Feynman diagrams for the $\Psi_a$, $\Pi_{ab}$ fields.

\subsection{2-point functions}

\begin{figure}[t]
\centering
\includegraphics[width=0.9\linewidth]{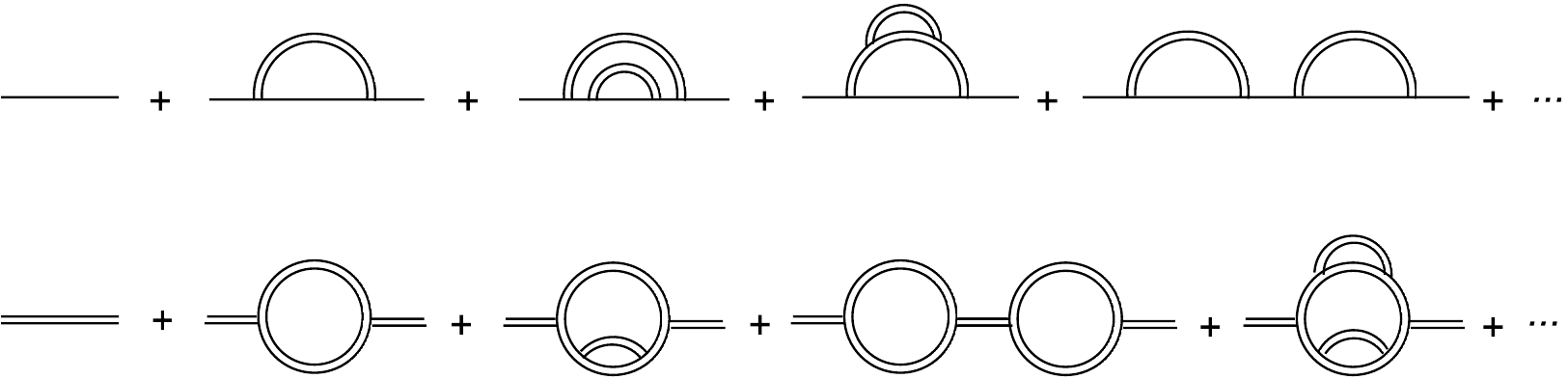}
\caption{The first few diagrams which dominate the $\psi\psi$ (top panel) and $\chi\chi$ (bottom panel) 2-point functions (and their super-descendants) in the large-$n$ limit when $g \sim 1/n$. In each panel the five diagrams scale as $g^k n^k$ for $k=0, 2, 4, 4, 4$, respectively. As noted in the text, single lines represent $\psi_a$ or $\b_a$ fields and double lines represent the mesonic $\chi_{ab}$ or $\pi_{ab}$ fields. All dominant diagrams are obtained by iterating the building blocks shown in figure~\ref{fig:susy4ptwithbos}.}
\label{fig:sd1}
\end{figure}

\begin{figure}[t]
\centering
\includegraphics[width=0.35\linewidth]{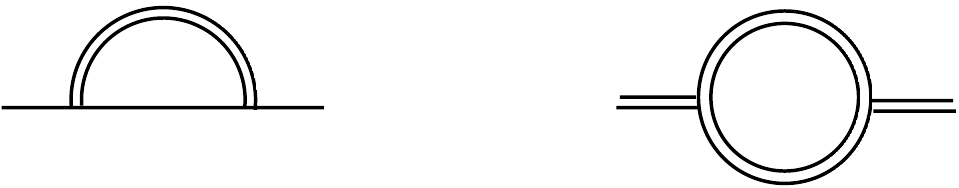}
\caption{Fundamental building blocks for the
$\psi\psi$ (left) and $\chi\chi$ (right)
2-point functions (and their super-descendants) at large $n$. Each of these two diagrams scales as $g^2n^2$.}
\label{fig:susy4ptwithbos}
\end{figure}

\begin{figure}[t]
\centering
\includegraphics[width=0.4\linewidth]{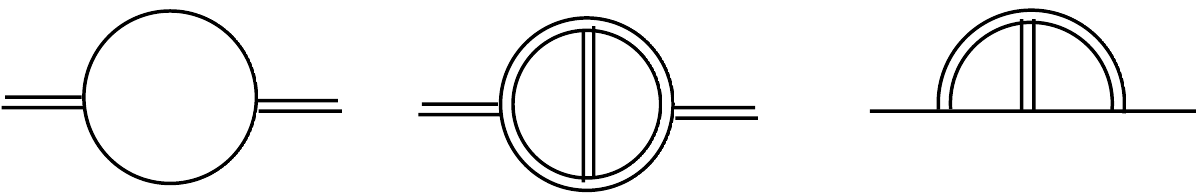}
\caption{Three examples of diagrams that are suppressed in the large-$n$ limit when $g \sim 1/n$. From left to right they scale as $g^2n$, $g^4n^2$ and $g^4n^3$ respectively.}
\label{fig:susy4ptwithbossubleading}
\end{figure}

\begin{figure}[t]
\centering
\includegraphics[width=0.65\linewidth]{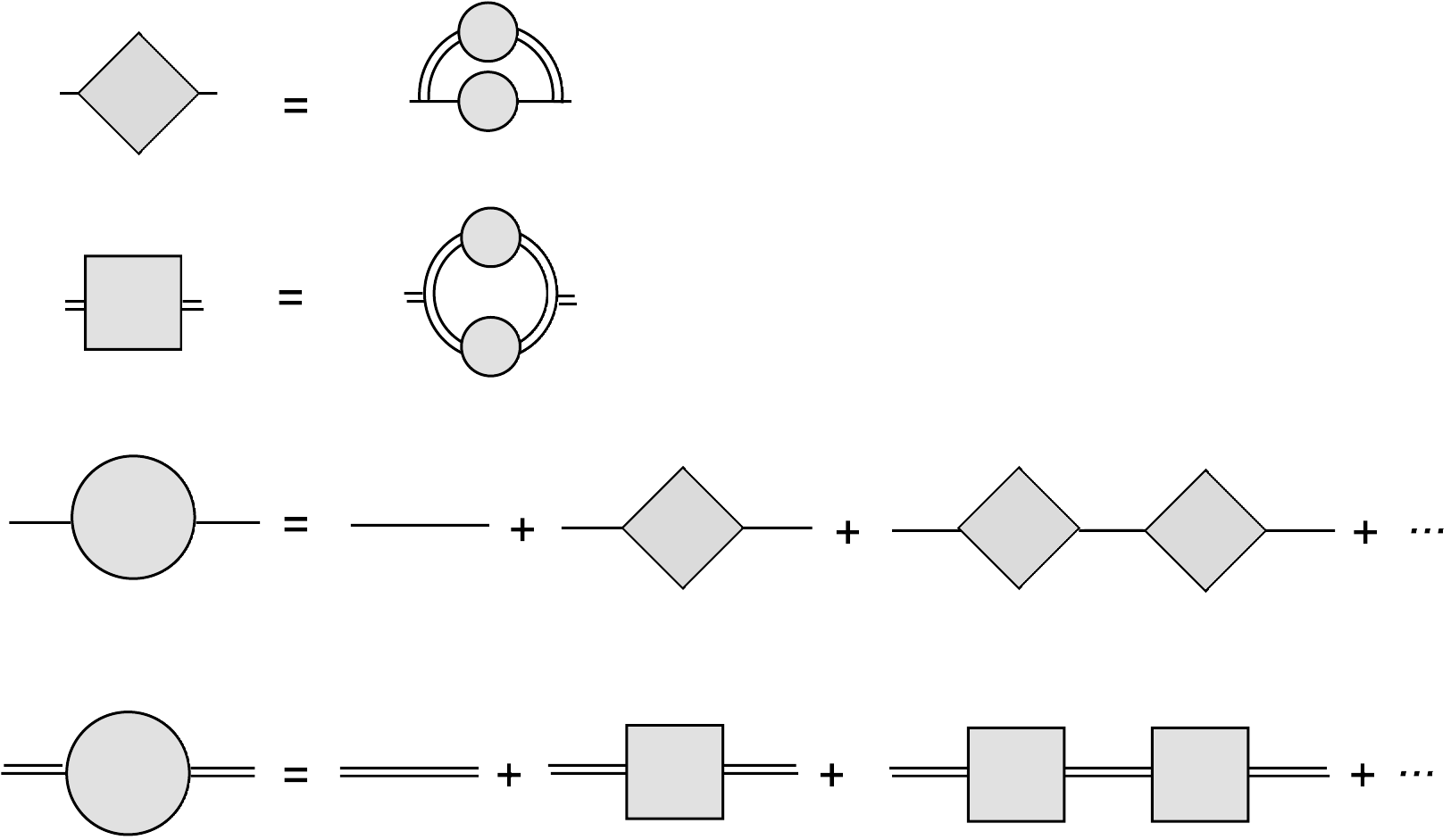}
\caption{Diagrammatic presentation of the Schwinger-Dyson equations. The diamonds represent the self-energy of the $\psi_a$ or $\b_a$ fields. The squares represent the self-energy of the $\c_{ab}$ or $\p_{ab}$ fields. The propagators with circular blobs are exact and are computed as shown in the bottom two panels.}
\label{fig:SD2}
\end{figure}

We show in the appendix that the dominant contributions to all 2-point functions scale as $g^{2k} n^{2k}$ in the large-$n$ limit.
Consequently, the large-$n$ limit exists if $g \sim 1/n$, as already advertised above. Several examples of these dominant contributions are shown in figure~\ref{fig:sd1}.
For completeness we also give a few examples of subdominant diagrams in figure~\ref{fig:susy4ptwithbossubleading}. All dominant contributions can be constructed by iterating the ``core'' building blocks shown in figure~\ref{fig:susy4ptwithbos}.

The iteration of the dominant large-$n$ contributions is represented diagrammatically in figure~\ref{fig:SD2}. Let us use $\Sigma^{xx}(t_1,t_2)$ to denote the self-energy of a field $x$, and $G^{xy}(t_1,t_2)$ to denote the 2-point function between a field $x(t_1)$ and another field $y(t_2)$. Then we can translate figure~\ref{fig:SD2} directly into a set of Schwinger-Dyson equations for the ``diagonal'' 2-point functions of the component fields:
\bal
\S^{\psi\psi}(t_1,t_2)&=-12{g^2}\, G^{\psi\psi}(t_1,t_2)G^{\pi\pi}(t_1,t_2)-12 {g^2}\, G^{\b\b}(t_1,t_2)G^{\c\c}(t_1,t_2)\,,\label{sd1}\\
\S^{\b\b}(t_1,t_2)&=-12{g^2}\, G^{\psi\psi}(t_1,t_2)G^{\c\c}(t_1,t_2)\,,\label{sd2}\\
\S^{\c\c}(t_1,t_2)&=-32{g^2}\, G^{\c \c}(t_1,t_2)G^{\pi\pi}(t_1,t_2)\,,\label{sd3}\\
\S^{\p\p}(t_1,t_2)&=-16{g^2}\, G^{\c\c}(t_1,t_2)^2\,,\label{sd4}
\eal
together with
\bal
\pa_{t_1} G^{\psi\psi}(t_1,t_3)+i\int dt_2\, \S^{\psi\psi}(t_1,t_2) G^{\psi\psi}(t_2,t_3)&=\delta(t_1-t_3)\,,\label{sd5}\\
\pa_{t_1} G^{\c\c}(t_1,t_3)+i\int dt_2\, \S^{\c\c}(t_1,t_2) G^{\c\c}(t_2,t_3)&=\delta(t_1-t_3)\,,\label{sd6}\\
iG^{\b\b}(t_1,t_3)+i\int dt_2\, \S^{\b\b}(t_1,t_2) G^{\b\b}(t_2,t_3)&=\delta(t_1-t_3)\,,\label{sd7}\\
iG^{\p\p}(t_1,t_3)+i\int dt_2\, \S^{\p\p}(t_1,t_2) G^{\p\p}(t_2,t_3)&=\delta(t_1-t_3)\,.\label{sd8}
\eal
The factors of $12$ in~\eqref{sd1} and~\eqref{sd2} are due to the 12 possible meson-fermion loops for a given $\psi_a$ or $\b_a$ 2-point function. The extra factors of  $32=8 \times 2^2$ in~\eqref{sd3} and~\eqref{sd4} are due to the 8 possible meson loops and the factor of 2 in the coupling (contributing a factor of $2^2$) in the last term of the Lagrangian~\eqref{Lcmp}. Finally, the  factor $16$ in~\eqref{sd4} is from the 4 possible meson loops, which is a half of the factor in~\eqref{sd3} because both the lines in the loop are $\c$ lines (in other words, we have multiplied by the familiar symmetry factor $\frac{1}{2}$).

This set of Schwinger-Dyson equations can be solved analytically in the IR limit, where the first term in each of the equations~\eqref{sd5}--\eqref{sd8} can be dropped. We make a power law ansatz for each field
\bal
G^{xx}(t_1,t_2)&\sim \frac{1}{(t_1-t_2)^{2\Delta_{x}}}\,, \qquad x \in \{\psi\,, \b\,,\p\,,\chi\}\,. \label{azt}
\eal
Plugging these back to the IR limit of eqs.~(\ref{sd1})--(\ref{sd8}) and carrying out the integrals, we get the following set of equations by a simple comparison of the powers of $(t_1-t_3)$ in the integration result:
\bal
\eqref{sd5} \Rightarrow \quad & 2\Delta_{\psi}+\Delta_{\p}=1\,,\qquad \Delta_{\psi}+\Delta_{\b}+\Delta_{\c}=1\,,\label{del1}\\
\eqref{sd6} \Rightarrow \quad & \Delta_{\p}+2\Delta_{\c}=1\,,\\
\eqref{sd7} \Rightarrow \quad & \Delta_{\b}+\Delta_{\psi}+\Delta_{\c}=1\,,\\
\eqref{sd8} \Rightarrow \quad & \Delta_{\p}+2\Delta_{\c}=1\,.\label{del4}
\eal
To completely determine the dimensions we follow~\cite{Fu:2016vas} in making use of supersymmetry. To that end let us first note that the tensorial analogue of eq.~(\ref{Qvar}) is clearly
\begin{align}
Q \psi_a(t)&=\b_a(t)\,, &  Q \b_a(t) &=i\pa_t \psi_a(t)\label{tsusy1}\,, \\
Q \c_{ab}(t)&=\p_{ab}(t)\,,& Q \c_{ab}(t) &=i\pa_t \c_{ab}(t)\,.
\label{tsusy2}
\end{align}
Indeed, as a consistency check, it is easy to verify that the Lagrangian~\eqref{Lcmp} changes by a total derivative
\begin{equation}
\delta_\x \cl=\frac{i\x}{2}\pa_t \big(-\psi_a\b_a-\c_{ab}{\Pi}_{ab}-\frac{g}{2}({\c}_{ab}\psi_a\psi_b+\frac{1}{2}\e_{abcd}\c_{ab}\psi_c\psi_d)+\frac{2g}{3} \c_{ab}\c_{bc}\c_{ca}\big)
\end{equation}
under the supersymmetry transformations~(\ref{tsusy1}) and~(\ref{tsusy2}). We also work out the expression of the supercharge in terms of the fundamental fields of this model, which is
\begin{equation}
Q= -\frac{\sqrt{2}g}{4}({\c}_{ab}\psi_a\psi_b+\frac{1}{2}\e_{abcd}\c_{ab}\psi_c\psi_d)+\frac{\sqrt{2}g}{3} \c_{ab}\c_{bc}\c_{ca}\,,
\end{equation}
which is normalized so that it squares to the Hamiltonian corresponding to~\eqref{Lsimple}:
\begin{equation}
Q^2=   \frac{g^2}{8} \e_{abcd}\psi_a\psi_b\psi_c\psi_d+\frac{g^2}{4} (\psi_a\psi_b)(\psi_a \psi_b)  -  2 g^2  \c_{ab} \c_{bc}\c_{cd}\c_{da}\,.
\end{equation}
Now if we apply these transformations to the 2-point correlation functions, in a manner identical to the derivation of eq.~(2.17) of~\cite{Fu:2016vas}, we obtain the relations
\bal
G^{\p\p}(t_1,t_2)&=-i\pa_{t_1} G^{\c\c}(t_1,t_2)\label{gg1}\,,\\
G^{\b\b}(t_1,t_2)&=-i\pa_{t_1} G^{\psi\psi}(t_1,t_2)\,.\label{gg2}
\eal
Together with eq.~(\ref{azt}) this implies
\begin{equation}
\Delta_{\b}= \Delta_{\psi}+\frac{1}{2}\,,\qquad \Delta_{\p}= \Delta_{\c}+\frac{1}{2}\,. \label{susyres}
\end{equation}
Now we have enough information to finally determine all of the IR scaling dimensions:
\begin{equation}
\Delta_\c=\Delta_\psi=\frac{1}{6}\,,\qquad \Delta_\p=\Delta_\b=\frac{2}{3}\,.\label{irresult}
\end{equation}

We can derive an additional consistency check on our large-$n$ Schwinger-Dyson equations by noting that eqs.~(\ref{gg1}) and~(\ref{gg2}) impose extra relations between the self-energies. Specifically, by plugging these into eqs.~(\ref{sd5})--(\ref{sd8}) and then integrating by parts we obtain
\bal
\S^{\psi\psi}(t_1,t_2)&=-i\pa_{t_1}\S^{\b\b}(t_1,t_2)\,,\label{sig1}\\
\S^{\c\c}(t_1,t_2)&=-i\pa_{t_1}\S^{\p\p}(t_1,t_2)\label{sig2}\,.
\eal
These conditions together with eqs.~(\ref{gg1}) and~(\ref{gg2}) are easily seen to be compatible with the Schwinger-Dyson equations~(\ref{sd1})--(\ref{sd8}).

In fact we can also easily determine the normalization of the propagators~\eqref{azt} in the IR limit. Plugging the ansatz
\bal
G^{xx}(t_1,t_2)&= \frac{n_x}{(t_1-t_2)^{2\Delta_{x}}}\,, \qquad x \in \{\b\,,\p\}\,, \label{azt1}\\
G^{xx}(t_1,t_2)&= \frac{n_x \,\text{sgn}(t_1-t_2)}{(t_1-t_2)^{2\Delta_{x}}}\,, \qquad x \in \{\psi\,,\chi\}\label{azt2}
\eal
into the IR limit of equations~\eqref{sd1}--\eqref{sd8} and using $t_{ij} \equiv t_i - t_j$ we obtain
\bal
-12i g^2 \int dt_2\, \Big(\frac{n_\psi n_\p\, \text{sgn}(t_{12})}{t_{12}^{2\Delta_{\psi}+2\Delta_{\p}}}+\frac{n_\c n_\b \, \text{sgn}(t_{12})}{t_{12}^{2\Delta_{\c}+2\Delta_{\b}}}\Big) \frac{n_\psi \,\text{sgn}(t_{23})}{t_{23}^{2\Delta_{\psi}}}=\delta(t_{13})\,,\label{irsd5}\\
-32i g^2 \int dt_2\,  \frac{n_\c n_\p\, \text{sgn}(t_{12})}{t_{12}^{2\Delta_{\c}+2\Delta_{\p}}}  \frac{n_\c \,\text{sgn}(t_{23})}{t_{23}^{2\Delta_{\c}}}=\delta(t_{13})\,,\label{irsd6}\\
-12 i g^2 \int dt_2\, \frac{n_\c n_\psi\, }{t_{12}^{2\Delta_{\c}+2\Delta_{\psi}}} \frac{n_\b}{t_{23}^{2\Delta_{\b}}}=\delta(t_{13})\,,\label{irsd7}\\
-16 i g^2 \int dt_2\, \frac{n_\c^2}{t_{12}^{4\Delta_{\c}}} \frac{n_\p}{t_{23}^{2\Delta_{\p }}}=\delta(t_{13})\,.\label{irsd8}
\eal
To solve the above set of equations it is useful to note the relations
\bal
\int dt_{13}\, e^{-i v t_{13}}\int dt_2\,   \frac{\text{sgn}(t_{12})}{|t_{12}|^\alpha}\frac{\text{sgn}(t_{23})}{|t_{23}|^{\beta}}&=  \frac{(2\p)^2}{c_f(\frac{1-\alpha}{2}) c_f(\frac{1-\beta}{2}) }|v|^{\alpha+\beta-2}\,,\\
\int dt_{13}\, e^{-i v t_{13}}\int dt_2\,  \frac{1}{|t_{12}|^\alpha}\frac{1}{|t_{23}|^{\beta}}&=  \frac{(2\p)^2}{c_b(\frac{1-\alpha}{2}) c_b(\frac{1-\beta}{2}) }|v|^{\alpha+\beta-2}\,,
\eal
which may be derived from the identities~\cite{Fu:2016vas}
\bal
\int dt\, e^{i w t} \frac{\text{sgn}(t)}{|t|^{2\Delta}}&=c_f(\Delta)  \frac{\text{sgn}(w)}{|w|^{1-2\Delta}}\,,\quad c_f(\Delta)=2i\, \cos(\p\Delta) \Gamma(1-2\Delta)\,,\\
\int dt\, e^{i w t} \frac{1}{|t|^{2\Delta}}&=c_b(\Delta)  \frac{1}{|w|^{1-2\Delta}}\,,\quad c_b(\Delta)=2\, \sin(\p\Delta) \Gamma(1-2\Delta)\ .
\eal
The equations~\eqref{irsd5}--\eqref{irsd8} then become, in the frequency domain,
\bal
\nonumber -12i g^2 \Big( \frac{n_\psi^2 n_\p\,(2\p)^2}{c_f(\frac{1-2\Delta_{\psi}-2\Delta_{\p}}{2}) c_f(\frac{1-2\Delta_{\psi}}{2}) }|v|^{2\Delta_{\psi}+2\Delta_{\p}+2\Delta_{\psi}-2}\qquad\qquad\qquad\qquad\\
\qquad \qquad\qquad+ \frac{n_\c n_\b n_\psi (2\p)^2}{c_f(\frac{1-2\Delta_{\c}-2\Delta_{\b}}{2}) c_f(\frac{1-2\Delta_{\psi}}{2}) }|v|^{2\Delta_{\c}+2\Delta_{\b}+2\Delta_{\psi}-2}\Big) =1\,,\label{irsd9}\\
-32i g^2 n_\c^2\, n_\p \frac{(2\p)^2}{c_f(\frac{1-2\Delta_{\c}-2\Delta_{\p}}{2}) c_f(\frac{1-2\Delta_{\c}}{2}) }|v|^{2\Delta_{\c}+2\Delta_{\p}+2\Delta_{\c}-2} =1\,,\label{irsd10}\\
-12 i g^2 n_\c n_\psi\,n_\b   \frac{(2\p)^2}{c_b(\frac{1-2\Delta_{\c}-2\Delta_{\psi}}{2}) c_b(\frac{1-2\Delta_{\b}}{2}) }|v|^{2\Delta_{\c}+2\Delta_{\psi}+2\Delta_{\b}-2}
=1\,,\label{irsd11}\\
-16 i g^2 n_\c^2 \, n_\p \frac{(2\p)^2}{c_b(\frac{1-4\Delta_{\c}}{2}) c_b(\frac{1-2\Delta_{\p }}{2}) }|v|^{4\Delta_{\c}+2\Delta_{\p }-2} =1\,.\label{irsd12}
\eal
Comparing the powers of $v$ leads to the relations~\eqref{del1}--\eqref{del4}:
\bal
\Delta_\c=\Delta_\psi\,,\qquad \Delta_\p=\Delta_\beta\,,\qquad \Delta_\p+2 \Delta_\c=1\ .\label{del5}
\eal
Taking the ratio of eqs.~\eqref{irsd10} and~\eqref{irsd12} gives
\bal
2 c_b(\textstyle{\frac{1-4\Delta_{\c}}{2}}) c_b(\textstyle{\frac{1-2\Delta_{\p }}{2}}) &=c_f(\textstyle{\frac{1-2\Delta_{\c}-2\Delta_{\p}}{2}}) c_f(\textstyle{\frac{1-2\Delta_{\c}}{2}}) \ .
\eal
Then making use of eq.~\eqref{del5} and the relations
\bal
c_f(\Delta)c_f(-\Delta)=-4 \pi \Delta \cot(\pi \Delta)\,, \qquad
c_b(\Delta)c_b(-\Delta)=-4 \pi \Delta \tan(\pi \Delta)
\eal
we arrive at
\begin{equation}
-4 \pi (1 - 4 \Delta_\chi) \cot(2 \pi \Delta_\chi) = - 2 \pi (1 - 2 \Delta_\chi) \tan(\pi \Delta_\chi)\,.
\end{equation}
This transcendental equation has infinitely many solutions,
but only one, $\Delta_\chi = 1/6$, that is consistent with
both eq.~(\ref{del5}) and the supersymmetry
constraint~\eqref{susyres}.
We will focus only on this solution in the following. The normalization factors $n_x $ in eqs.~\eqref{azt1} and~\eqref{azt2} can also be fixed once we take into account eqs.~\eqref{gg1} and~\eqref{gg2}, which give the relations
\bal
n_{\p}=2 i \Delta_\c n_\c\,, \qquad n_\b=2 i \Delta_\psi n_{\psi}\ .
\eal
Finally we conclude that the equations~\eqref{irsd9}--\eqref{irsd12} are solved by
\bal
n_\c=- (32\sqrt{3} \pi g^2)^{-1/3}\,,\qquad n_\psi=\pm (36 \pi g^2)^{-1/3}\ .
\eal

Let us make two comments on eq.~(\ref{irresult}). First, even though the meson fields dominate in the large-$n$ limit defined by the scaling~(\ref{gn}), the original $\psi_a$ and $\b_a$ fields experience similar IR physics, having exactly the same IR dimensions as the $\c_{ab}$ and the $\p_{ab}$ fields, respectively. This can be understood as a consequence of supersymmetry, which introduces non-dynamical bosonic fields that in turn provide the constraints~(\ref{eq:constraint}) relating the pair $(\psi_{a},\b_{a})$ with $(\c_{ab},\p_{ab})$. Second, it is notable that the IR dimensions $\frac{1}{6}, \frac{2}{3}$ of our tensor fields are exactly the same as those of the component fields in the supersymmetric FGMS model of~\cite{Fu:2016vas}. This provides encouraging support to our supersymmetrization approach; since the fermionic tensor model of~\cite{Witten:2016iux} has the same IR physics as the SYK model (in the large-$n$ limit), our supersymmetric tensor model should give the same IR physics as that of the supersymmetric FGMS model introduced in~\cite{Fu:2016vas} if our supersymmetrization is consistent.

\subsection{4-point functions}

\begin{figure}[t]
\centering
\includegraphics[width=0.20\linewidth]{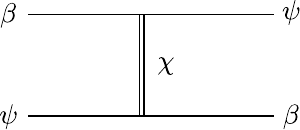}
\caption{The leading large-$n$ contributions to the
$ \psi \psi \beta \beta$ 4-point functions are obtained
by iterating this kernel, leading to ladder diagrams of the
type shown in the last panel of figure~\ref{fig:4ptgen}.}
\label{fig:kernel}
\end{figure}

We can also consider 4-point functions. In the appendix we show that the dominant large-$n$ contributions to 4-point functions of the $\psi_a, \beta_a$ fields come from meson-exchange ladder diagrams of the type shown in the last panel of figure~\ref{fig:4ptgen}. The simplest example is a 4-point function of the type $\<\psi_a(t_1)\b_b(t_2)\psi_a(t_3)\b_b(t_4)\>$. Plugging in the appropriate component fields, we find that contributions to this 4-point function are obtained by iterating the kernel shown in figure~\ref{fig:kernel}. This kernel has essentially the same form as the analogous one in the supersymmetric FGMS model (see figure 6(a) of~\cite{Fu:2016vas}). Moreover, since we found in the previous subsection that the IR dimensions of our fields coincide precisely with those of that reference, we conclude that the 4-point functions of our supersymmetric tensor model must be the same as those of those in the supersymmetric FGMS model of~\cite{Fu:2016vas}. In particular, this means that the eigenvalues of the kernel and hence the operator spectrum that appears in the OPE limit of the 4-point functions must also be the same as those in~\cite{Fu:2016vas}.

\section{Discussion}\label{diss}

The Gurau-Witten tensor model of~\cite{Witten:2016iux}
is based on an interaction
of the form
\begin{equation}
\cl_{\rm int} = j\, \epsilon_{abcd} \psi_a \psi_b
\psi_c \psi_d
\label{eq:int1}
\end{equation}
and has a well-defined large-$n$ limit if $j \sim n^{-3/2}$, in which
case it reproduces exactly the physics of the large-$n$ limit of the fermionic SYK model.

In this paper we introduced a tensor model that has
the same IR physics as the supersymmetric FGMS
model of~\cite{Fu:2016vas} in the large-$n$ limit.
Our construction required the introduction
of tensor-valued ``meson'' fields $\chi_{ab}$ as well as auxiliary boson
fields.
In our analysis we noted the fact that taking the large-$n$ limit
and integrating out the auxiliary bosons do not commute.
If we first take the large-$n$ limit with $g \sim 1/n$, we find that
the effective dynamics of the quark fields in the IR is
the same as that of the supersymmetric FGMS model.

However, if we first integrate out the bosonic fields, we found that
a judicious
choice of coupling constants decouples the mesons completely, leaving
an effective interaction
\begin{equation}
\label{eq:int2}
\cl_{\rm int} = - \frac{g^2}{8} \e_{abcd}\psi_a\psi_b\psi_c\psi_d - \frac{g^2}{4} (\psi_a\psi_b)(\psi_a\psi_b)
\end{equation}
which involves~\eqref{eq:int1}
as well as the only other operator that is quartic in $\psi$ fields
and invariant under the global symmetry group. However,
the large-$n$ limit of this model is different from Gurau-Witten; the presence of the pillow operator violates the
large-$n$ limit $g^2 \sim n^{-3/2}$ that is appropriate for the interaction \eqref{eq:witteninteraction}.\footnote{We thank Igor Klebanov and Grigory Tarnopolsky pointing out this important fact to us.}
It will be worthwhile to explore the large-$N$ property of this pillow operator in the future.

Further notice that we can consider a purely mesonic model, which can be obtained from \eqref{supertensorLhg} by turning off all the $\Psi_a$ fields.\footnote{We thank Juan Maldacena for raising this point to us.} Since the $\Psi_a$ fields are subdominant, we believe that this mesonic model has the same IR physics as we discussed in this paper. We include the $\Psi_a$ fields in order to make the comparison with the purely fermionic Gurau-Witten model manifest, although our model does not reduce to the latter.

\acknowledgments

We would like to thank Antal Jevicki, Jeff Murugan, Kenta Suzuki
for useful discussions, and are especially
grateful to Igor Klebanov, Juan Maldacena and Grigory Tarnopolsky for important comments on
the draft.
This work was supported by the US Department of Energy under contract DE-SC0010010 Task A and by Simons Investigator Award \#376208 (AV).

\appendix

\section{The mesonic melon patch: a proof of the large-$n$ expansion}\label{pf}

Here we prove that the large-$n$ limit of~(\ref{Lcmp}) is dominated by ``melonic'' graphs of the type shown in figure~\ref{fig:sd1}. The proof is split into several steps.
We first discuss vacuum graphs, since graphs with external
fields can be obtained by cutting lines in vacuum graphs.
Moreover we start with vacuum graphs with only mesonic
loops since quark loops are suppressed by $1/n$, as we show in step~5.

\subsection*{Step 1: A Convenient Notational Trick}

\begin{figure}[t]
\centering
\includegraphics[width=0.5\linewidth]{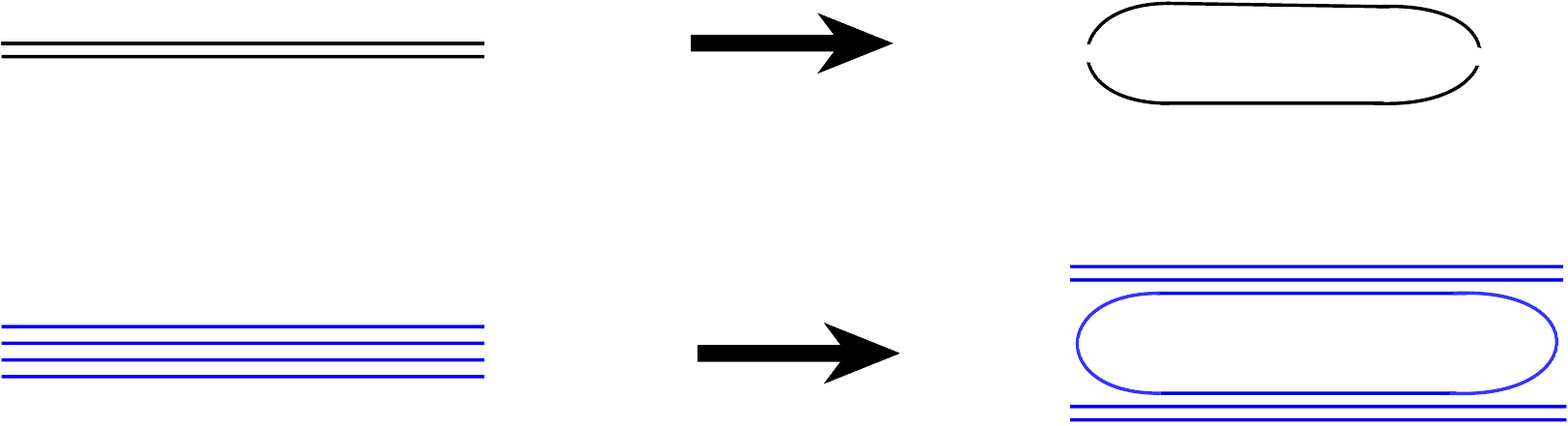}
\caption{Pulling apart a mesonic double line into a pair of single quark lines. The upper panel shows this process in terms of the single/double line notation introduced in figure~\ref{fig:sd1}; the lower panel depicts the same splitting in terms of the four individual $O(n)$ index strands carried by the meson field.}
\label{fig:split_prop}
\end{figure}

We start by introducing a modification to the way we draw our Feynman diagrams, such as those in figure~\ref{fig:sd1}, to manifest the connection to the fermionic model. This is done by ``pulling apart" each internal double line associated to a mesonic $\p_{ab}$ or $\c_{ab}$ field into a pair of single lines. In terms of the individual vector index strands (of the different $O(n)$ groups) to which the single and double lines resolve, this process inserts a strand loop of type $\overline{ab}$ into each internal mesonic line of type $ab$. This process, depicted in figure~\ref{fig:split_prop}, allows us to translate any diagram in our model into one with only three-stranded internal lines. By redrawing every diagram in this way we will be able to transcribe many steps of the large-$n$ proof presented in~\cite{Witten:2016iux} (and based on~\cite{Bonzom:2011zz}).

\begin{figure}[t]
\centering
\includegraphics[width=0.9\linewidth]{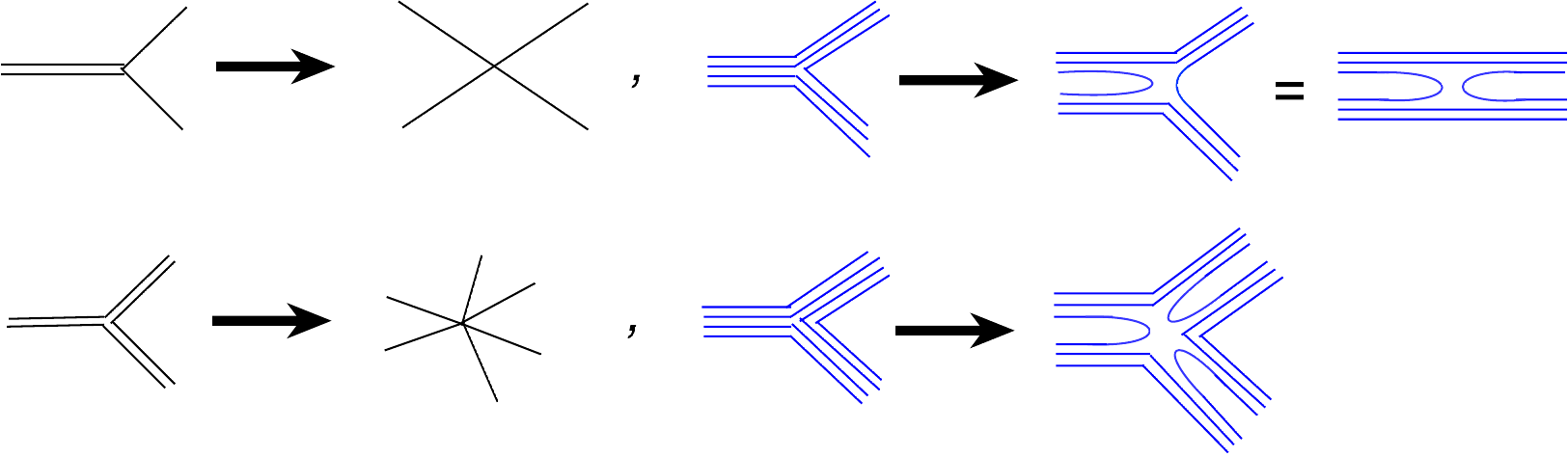}
\caption{Modification of the vertices. The top left panel shows the resolution of the $\pi \psi \psi$ vertex into a quartic vertex of single lines, and the bottom left panel shows the resolution of the $\chi \chi \chi$ vertex into a 6-valent vertex. The right panels show the same procedure in terms of the individual vector index strands.}
\label{fig:split_vertex}
\end{figure}

Notice that this procedure is only defined for internal double lines in a diagram, which always end on two vertices.
Consequently we should modify the vertices as well. The original $\pi_{ab} \psi_a \psi_b$ vertex (and its superpartner $\chi_{ab} \beta_a \psi_b$) splits into a quartic vertex for single lines, and the original $\pi_{ab} \chi_{bc} \chi_{ca}$ vertex splits into a 6-valent vertex. When these vertices are resolved into the individual index strands they reveal that the extra strand loop inserted into each meson line undergoes a ``U-turn'' at every vertex. These modifications are pictured in figure~\ref{fig:split_vertex}.

To summarize, each vacuum diagram in the theory~(\ref{Lcmp}) can be drawn with single-line three-stranded edges interacting via 4-valent and 6-valent vertices. The benefit of this rewriting is that we can basically transcribe the large-$n$ counting argument from~\cite{Witten:2016iux,Bonzom:2011zz}. The only change we have to make is to remove one power of $n$ for every extra loop of strand that was introduced, i.e.~one power of $n$ for every meson line in the original diagram.

\subsection*{Step 2: Counting the Large-$n$ Scaling of a Graph}

At this step we are focusing on vacuum diagrams with only 6-valent vertices. Proceeding as in~\cite{Witten:2016iux,Bonzom:2011zz} we consider one at a time the three separate ways to ``project'' a stranded diagram onto a standard double-line type Feynman diagram. Here by ``double-line'' we mean fat diagrams of the type familiar in the study of large-$N$ gauge theory and not the double lines we have been using to denote our meson fields; importantly, there is no meaning of planarity for the latter. To define a projection we first pick one of the three cyclic orderings of the labels $(0,1,2,3)$ modulo reflection, i.e.
\begin{equation}
\label{eq:jchoices}
\cj =
(0,1,2,3), (0,1,3,2), ~ \text{or} ~(0,2,1,3)\,.
\end{equation}
Then, for a given ordering $\cj=(\ldots, a_{i-1} , a_i, a_{i+1},\ldots)$, we keep only the two vector index strands $\overline{a_{i}a_{i\pm 1}}$ of each single line of type $a_i$, and we ignore the third index strand $a_{i+2} = a_{i-2}$. Then we glue a ``face'' of type $\cf_{a_i,a_{i+1}}$ onto each closed index loop of type $\overline{a_{i}a_{i+ 1}}$. In this way each diagram becomes a fat line diagram as in conventional large-$N$ theories. We denote the collection of all faces for a given $\cj$ by
\begin{equation}
\cf^\cj=\sum_{i=0}^3 \cf^\cj_{a_ia_{i+1}}\,.
\end{equation}
The Euler characteristic of the resulting fat diagram is, for a given $\cj$,
\begin{equation}
\c_\cj=V_6-E+\cf_\cj=
-2V_6+\sum_i \cf^\cj_{a_ia_{i+1}}\,,
\end{equation}
where the $V_6$ is the number of 6-valent vertices and $E=( 6V_6)/2$ is the total number of edges. Notice that $E$ is not the total number of lines in the original graph before splitting open all of the mesonic lines; instead, the number of mesonic lines in the original graph is
\begin{equation}
e_2=(3V_6)/2\,. \label{e2}
\end{equation}
We define the ``degree" of a given single-line graph by
\begin{equation}
\w=\sum_\cj \left(1-\frac{\c_\cj}{2}\right),
\end{equation}
which evaluates to
\begin{equation}
\w=3+3V_6-\cf\,.\label{cjh}
\end{equation}
Here $\cf=\sum_{i} \cf_{a_ia_{i+1}}$ is the total number of closed loops of strands in the graph; in particular, this should not be confused with the $\cf^\cj$ that is defined for a given fat diagram corresponding to the ordering $\cj$.

Given these relations, each graph scales with $g$ and $n$ according to
\begin{equation}
g^{V_6} n^{\cf}n^{-e_2}\,,
\end{equation}
where the last factor removes the extra loops we introduced in step~1 to inflate the original mesonic lines. Using~\eqref{gn}, \eqref{e2} and~\eqref{cjh}, this evaluates to
\begin{equation}
n^{3+\tfrac{1}{2}V_6-\w}\,.\label{npower}
\end{equation}
It may look worrisome that the result depends on the number of 6-valent vertices with a positive coefficient, since it might allow for arbitrarily large $n$-scaling for diagrams with sufficiently many 6-valent vertices.
However, we will show in the next step that
\begin{equation}
\label{eq:bound}
\frac{1}{2} V_6 - \omega \le 1 ~ \text{for all graphs.}
\end{equation}

\begin{figure}[t]
\centering
\includegraphics[width=0.3\linewidth]{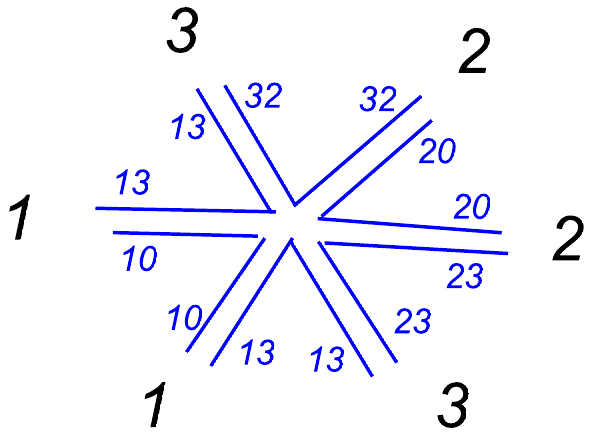}
\caption{A 6-valent vertex with index labels $\{ 1,1,2,2,3,3\}$. When we resolve this into a fat vertex for the choice $\cj = (0,1,3,2)$, it is necessary to arrange the two 1's to be cyclically adjacent, the two 2's to be cyclically adjacent, and the 1's and 2's to be separated by 3's. This is the only way to ensure that the strands of type $\overline{10}$ and $\overline{20}$ match up. For different choices of $\cj$ the vertex would have to be resolved into a fat diagram with different cyclic orderings.}
\label{fig:order}
\end{figure}

\subsection*{Step 3: Bounding the Large-$n$ Behavior}

To show~(\ref{eq:bound}) we take a closer look at the form of the 6-valent vertex. This vertex originates from the $\pi_{ab} \chi_{bc} \chi_{ca}$ coupling in eq.~(\ref{Lcmp}). Therefore, when we pull the meson lines apart into single lines as explained in step~1, this vertex resolves into a coupling of three pairs of distinct indices $\{ a,a,b,b,c,c\}$ and, importantly, does not couple arbitrary sets of indices.
The notation $\{ \cdots \}$ emphasizes that we are talking merely about a set of indices; there isn't yet any notion of cyclic ordering.

Now consider some given $\cj$ and suppose, without loss of generality, that the three labels $a$, $b$, $c$ in the 6-vertex appear in the cyclic order $\cj = (a, b, c, d)$, where $d$ is the fourth index, i.e.~$\{d\} = \{0,1,2,3\} \setminus \{a, b, c\}$.
Then, when we resolve into a fat diagram using this $\cj$,
the index $a$ resolves into strands of type
$\overline{ab}$ and $\overline{da}$,
the index $b$ resolves into strands of type
$\overline{ab}$ and $\overline{bc}$,
and the index $c$ resolves into strands of type $\overline{bc}$ and
$\overline{cd}$. In total we have
12 strands. In order for the two $\overline{da}$ strands to pair
up, and for the two $\overline{ab}$ strands to pair up, it is necessary
to order the 6 index labels of the vertex cyclically as
$(a,b,c,c,b,a)$, i.e., so that the two $c$'s are cyclically adjacent,
the two $a$'s are cyclically adjacent, and these pairs are separated
from each other by the two $b$'s.
An example of this is shown in figure~\ref{fig:order}
for the choice $\{a,b,c,d\} = \{1,3,2,0\}$.

\begin{figure}[t]
\centering
\includegraphics[width=1.0\linewidth]{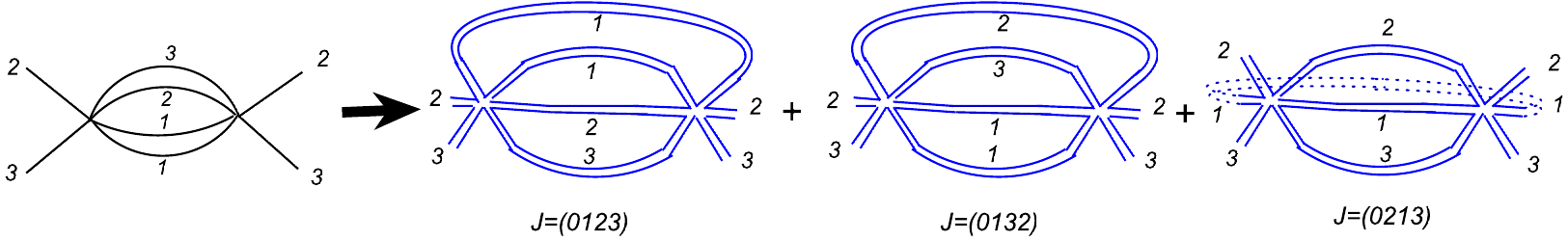}
\caption{An example illustrating the existence of a handle for one of the three $\cj$'s in the case when two 6-valent vertices are connected by four lines.
The 3 diagrams to the right of the arrow show the three blow-ups of this graph for the three different choices $\cj=(0123)$, $(0132)$ and $(0213)$ respectively. In this example we see that only the third ordering develops a handle.
Inserting a pair of 6-valent vertices of this type into a larger
diagram therefore
does not change the value of $\frac{1}{2}V_6 - \omega$ since
$V_6$ increases by two while $\omega$ increases by one.
}
\label{fig:handle}
\end{figure}

\begin{figure}[t]
\centering
\includegraphics[width=0.65\linewidth]{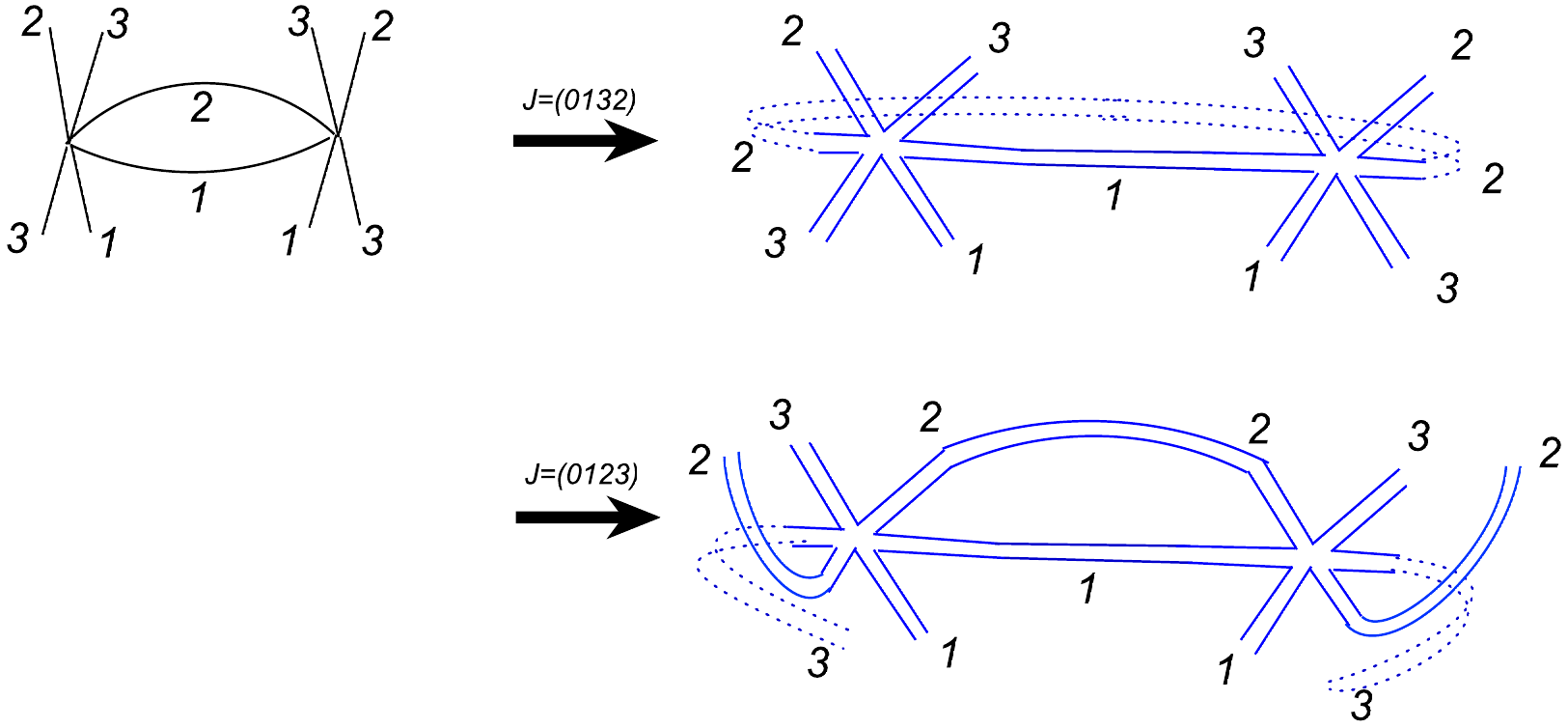}
\caption{In contrast to the situation shown in figure~\ref{fig:handle}, if two 6-valent vertices are connected by only two lines, then the corresponding fat graph is non-planar for two $\cj$'s.
Inserting a pair of 6-valent vertices of this type into a larger graph
therefore
decreases $\frac{1}{2}V_6 - \omega$ by one.}
\label{fig:handle_wrong}
\end{figure}

There are three possible $\cj$'s (see eq.~(\ref{eq:jchoices})), so there are
three different ways to blow up a given 6-valent vertex into the
vertex in a fat line diagram.
It
is clear that the blow-up of a single 6-vertex
can always be drawn as a planar fat vertex.
However, as soon as two such vertices are joined,
in general there will be a topological obstruction to maintaining planarity.
Specifically, it is easy to see that whenever two 6-valent vertices
are joined together, there must be a handle
for at least one of the three $\cj$'s, corresponding to a twist
in matching the fat lines to the original legs.
More specifically, if the pair of 6-valent vertices is joined
by four lines (i.e., by two mesons, in the original diagram), then
a handle exists for only one $\cj$, as shown in figure~\ref{fig:handle}.
On the other hand, if the pair of vertices is joined by only
two lines (i.e., by one meson), then a handle must exist
for more than one $\cj$, as shown in figure~\ref{fig:handle_wrong}.

The only exception occurs when two 6-valent vertices are connected by 6 legs,
as shown in figure~\ref{fig:bubble}.
This can be blown up into a planar fat graph for each choice of $\cj$,
so this diagram has $V_6 = 2$, $\omega = 0$, and hence
$\frac{1}{2} V_6 - \omega = 1$ so it grows like $n^4$
according to eq.~(\ref{npower}). (The 1-loop diagram with
no vertices should be thought of as a special case, it clearly
scales as $n^4$ also.) We take the bubble graph of
figure~\ref{fig:bubble} as the starting point of our inductive construction
to be discussed in the next step.
Here we finish by noting that since adding a pair of 6-valent
vertices
(the number of such vertices must always be even)
to any
graph necessarily introduces a handle for at least one $\cj$,
so the quantity $\frac{1}{2} V_6 - \omega$ can never be greater than 1.
This completes the proof of eq.~(\ref{eq:bound}).

\subsection*{Step 4: Identifying the Dominant Vacuum Graphs}

\begin{figure}[t]
        \centering
        \includegraphics[width=0.1\linewidth]{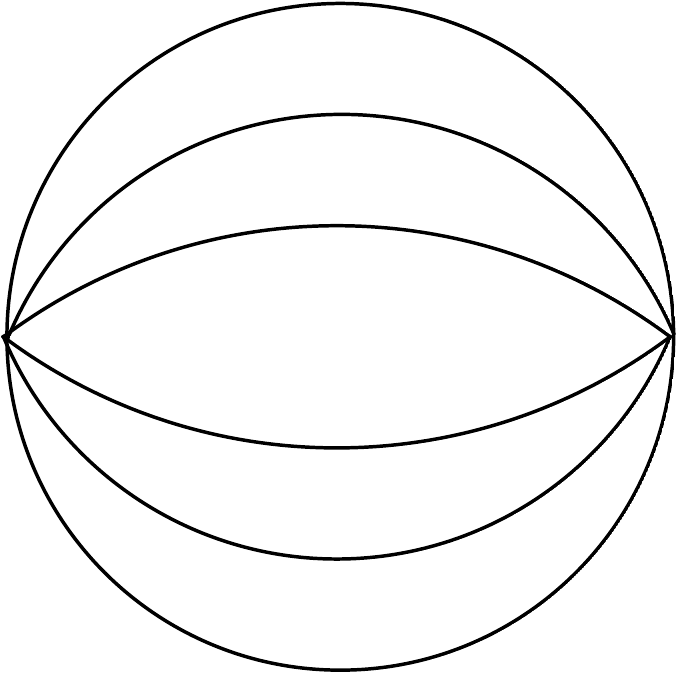}
        \caption{The vacuum bubble diagram, which is a leading contribution to the large-$n$ limit, having $V_6 = 2$ and $\omega = 0$. It contains two 6-valent vertices and can be blown up into a planar fat graph for each of the three choices of $\cj$. Other dominant diagrams can be recursively obtained by adding 6-valent vertices to this graph in a manner explained in step~4 of the appendix.}
        \label{fig:bubble}
\end{figure}

At this stage we have now proven that there is indeed a well-defined large-$n$ expansion when $g \sim 1/n$.
Our next task is to identify the vacuum graphs that dominate this limit.
These are the
ones which saturate the bound
\begin{equation}
\omega = \frac{1}{2} V_6 - 1\label{cond}
\end{equation}
and hence scale like $n^4$.
As in~\cite{Witten:2016iux,Bonzom:2011zz} we proceed by induction starting with the graph shown in figure~\ref{fig:bubble}. Now suppose we have a graph that satisfies~\eqref{cond}. Next we consider adding a pair of 6-valent vertices into this graph. The bound~(\ref{cond}) will continue to hold only if the addition of these vertices introduces a handle for a single $\cj$; if handles are introduced for more than one $\cj$ then~(\ref{cond}) will be violated.

\begin{figure}[t]
        \centering
        \includegraphics[width=0.6\linewidth]{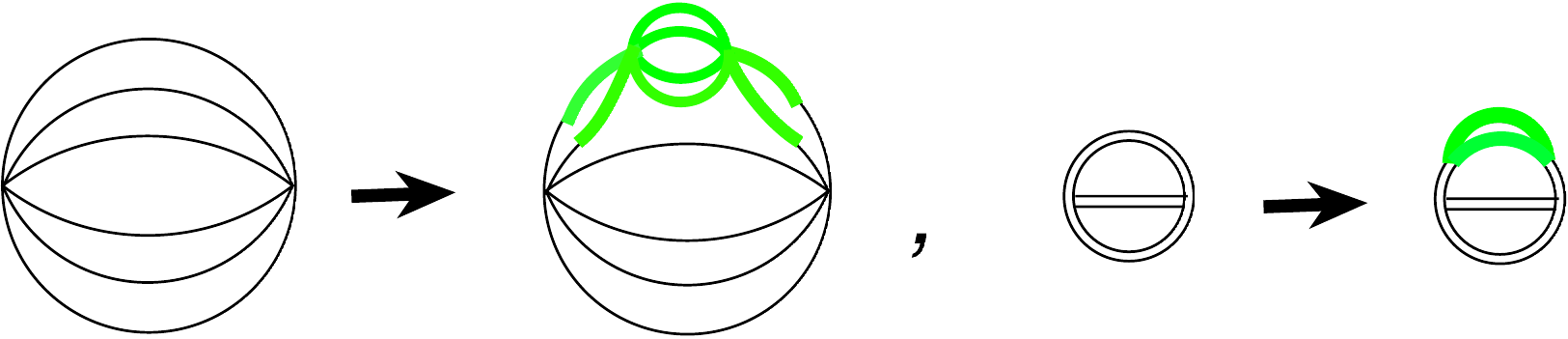}
        \caption{An example of the recursive generation of
dominant ($\mathcal{O}(n^4)$) vacuum diagrams. The process is
shown in terms of the inflated single line propagators and
vertices on the left; and in terms of the original double line
notation on the right. The green part in bold lines is the building
block shown in figure~\ref{fig:handle}. Iterating this procedure leads to ``melon'' graphs of the standard type.}
        \label{fig:recgen}
\end{figure}

Since each pair of adjacent vertices in a graph
must be connected by at least one
handle, it is clear that if we add two new vertices to a graph, the only
way to avoid adding more than one new handle is for there to be
one handle between the two new vertices.  If instead the new
handles connect new vertices to existing ones, then at least 2 new handles
are needed, which gives a subdominant diagram.  But the only way to
add a pair of 6-valent vertices connected by exactly one handle is to use
a pair of the type shown in figure~\ref{fig:handle}.

We conclude that the dominant diagrams are all built from the vacuum graph in figure~\ref{fig:bubble} by breaking some pairs of the inflated propagators and inserting the building block shown in figure~\ref{fig:handle} in all possible ``planar" ways.
One example of such an insertion is shown in figure~\ref{fig:recgen}. Each resulting diagram can then be translated back to the original double-line notation.

\subsection*{Step 5: The Dominant Contributions to Meson and Quark Correlators}

\begin{figure}[t]
\centering
\includegraphics[width=0.7\linewidth]{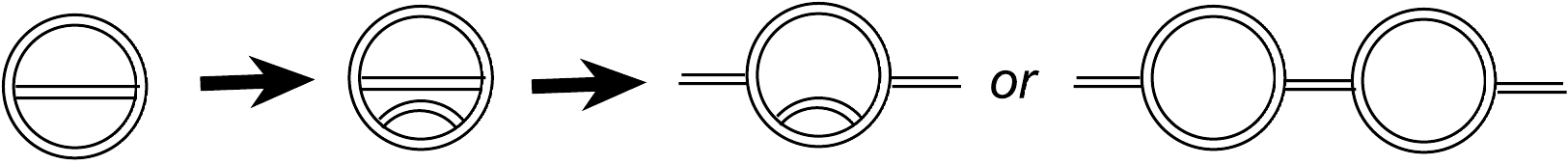}
\caption{An example of the recursive construction that generates the leading contributions to the 2-point function of the original double-line (meson) fields.
The first arrow represents the action of inserting one building block,
as in figure~\ref{fig:recgen}. The third panel shows two different ways
to get a 2-point graph by cutting one of the edges in the second panel.}
\label{fig:2pt_example_double}
\end{figure}

To find the dominant contributions to the meson propagator we can simply cut any edge in one of the dominant vacuum graphs.  This gives precisely the set of graphs shown already in~\ref{fig:sd1}.  One example of this cutting procedure is shown in figure~\ref{fig:2pt_example_double}.

Finally we are ready to construct the dominant contributions to the
correlation functions of the single-line quark fields.
These are always obtained by cutting vacuum graphs with a single quark
loop (and arbitrarily many meson loops), as we now discuss.

First, to get a vacuum graph with single lines we need to insert some
4-valent vertices of the type shown in the top panel
of figure~\ref{fig:split_vertex}.  It is clear from
the form of this vertex that it can be interpreted simply
as splitting a double (meson) line into a pair of single (quark)
lines.
This interaction originates from the vertices with one double line and two
single lines, i.e.~$\pi_{ab} \psi_a \psi_b$ or $\chi_{ab} \beta_a \psi_b$.
But since the two single lines in any such vertex necessarily carry
different indices $a \ne b$, they cannot close to form a tadpole that
we could cut to get a contribution to a quark two-point function.
To get such a contribution we need to insert at least two
4-valent vertices.

\begin{figure}[t]
\centering
\includegraphics[width=0.7\linewidth]{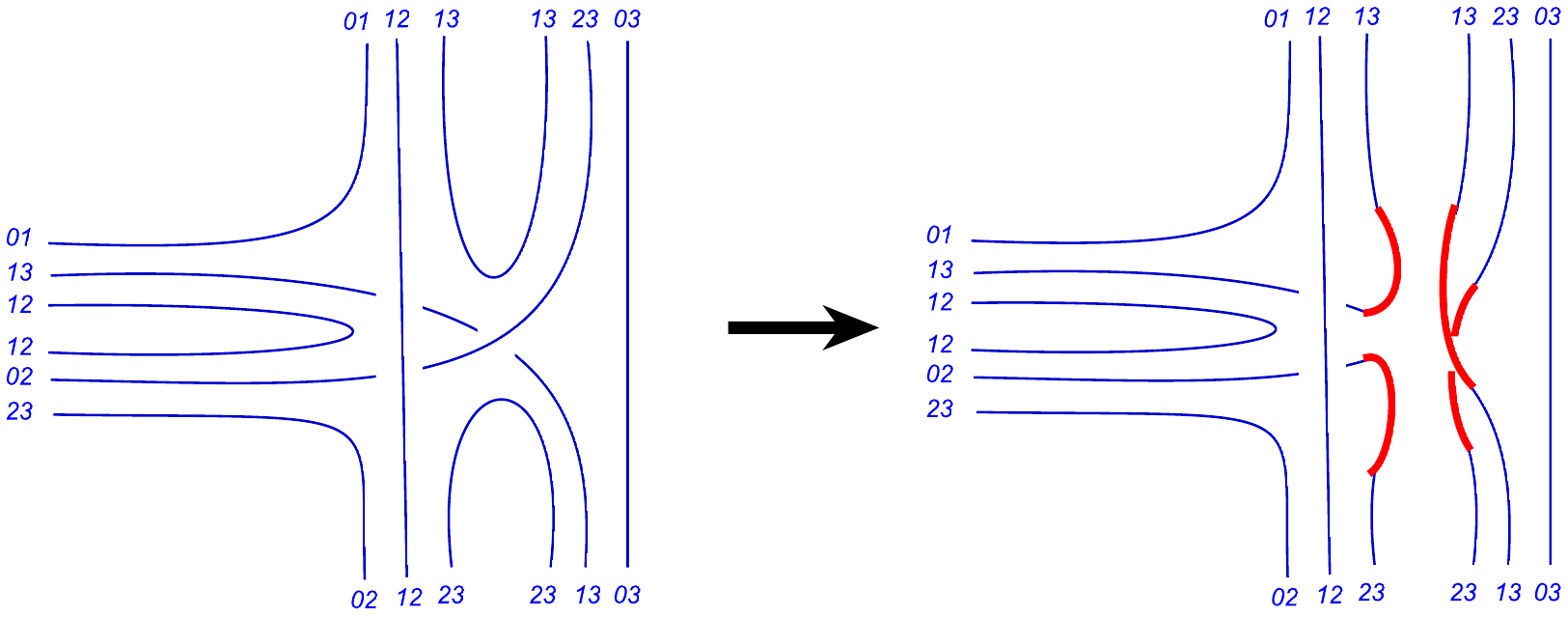}
\caption{The move to separate out a single line from a 6-valent
vertex to make a 4-valent vertex. In the left panel, the 6-valent
vertex carries indices $\{1,1,2,2,3,3\}$ and we can strip off the
line $3$ by breaking 2 virtual loops and reconnecting the open ends shown by the bold strands in red.}
\label{fig:6to4vertex}
\end{figure}

In addition to the insertion of these 4-valent vertices, we need one additional move in the following discussion, which is to strip off a line from a 6-valent vertex to make it a 4-valent vertex. This process breaks two of the virtual loops we introduced in step~1 and rejoins them after an exchange. An example of this move is shown in figure~\ref{fig:6to4vertex}. It is clear that this move does not change the actual number of index strand loops in the diagram. In addition, since the number of vertices also does not change, we conclude that this move does not change the large-$n$ behavior of a diagram. Finally we note that there is symmetry between the 3 pairs of indices of each 6-valent vertex, and we always have a free choice of which pair of indices to strip off.

\begin{figure}[t]
\centering
\includegraphics[width=0.8\linewidth]{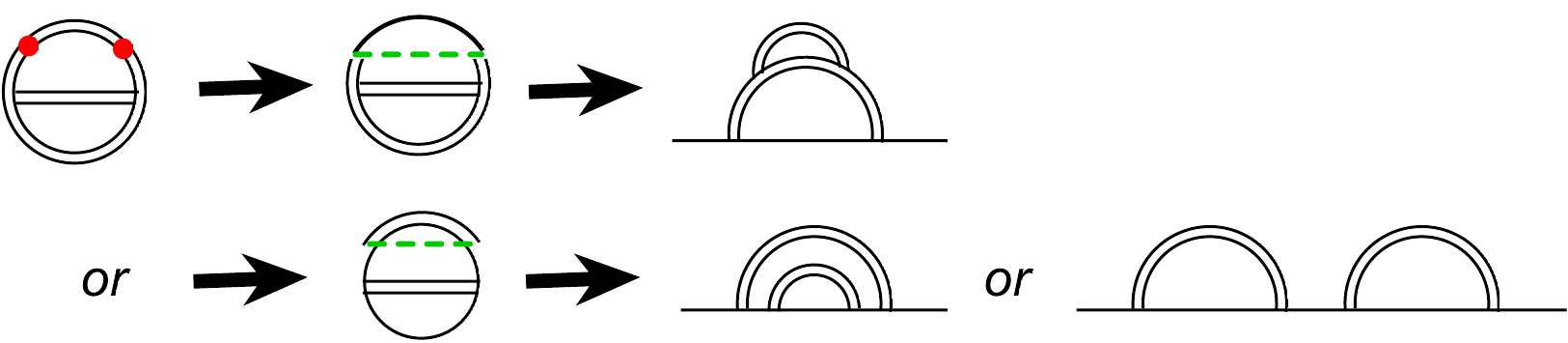}
\caption{An example showing how to generate dominant
contributions to the quark 2-point functions. We start with a
mesonic vacuum graph and insert two vertices, shown in red.
In this figure the two vertices sit on the same meson propagator,
but this needn't be the case in general.  We then ``unzip'' the
mesons between these vertices.  This can be done in two ways.
In the top panel we unzip the ``short'' way and in the bottom
panel we unzip the ``long'' way.  In the latter case, as
discussed in the text, the unzipping passes straight through
the two vertices. Finally we obtain various dominant
contributions to the quark 2-point function by cutting
open the various quark lines that have been exposed.}
\label{fig:2pt_example_single}
\end{figure}

Now consider the process of adding a single quark loop inside any
vacuum diagram of the meson fields.
We simply start with any dominant ``melon'' contribution to the vacuum (zero-point function). Then consider all possible distinct ways of inserting a pair of 4-valent vertices onto the various double-line propagators. The two vertices may be on the same propagator, or on different propagators.
Then, starting from one of these vertices, and proceeding in either direction,
we can ``unzip'' the meson double lines into a pair of quark lines.
Each single line opened up in this way extends until it hits the next
vertex.
If this is a 6-valent vertex, then we can use the move shown
in figure~\ref{fig:6to4vertex} to reduce the 6-valent vertex
into a 4-valent vertex with the single line cleanly separated out.
Then we simply continue unzipping past this vertex.
Ultimately, the zipper runs out when we hit the second of our two
inserted 4-valent vertices.
Note that for each pair of inserted vertices, there is more than one
distinct way to unzip a diagram.
An example of this process is shown in figure~\ref{fig:2pt_example_single}.

All dominant vacuum diagrams with a single quark loop are obtained
in this way by unzipping some double-line in a melonic vacuum graph.
How do these graphs scale with $n$?  As discussed above, the unzipping
move shown in figure~\ref{fig:6to4vertex} does not affect the scaling
with $n$.  We get a factor of $1/n^2$ from having inserted two 4-valent
vertices.
But this process also creates precisely one new strand loop, since
the two 3-strand
single lines from one of the insertions have to join up with the two 3-strand
single lines from the other insertion, giving a factor of $n$.
Altogether we conclude that the leading vacuum graphs with a single
quark loop scale as $n^3$ in the large-$n$ limit.
Of course,
it is expected that each quark loop should suppress a graph
by $1/n$ compared to a graph with only meson loops.

With all these preparations, we finally can construct the dominant diagrams contributing to the 2-point functions of some single-line fields simply by
cutting open the quark loop.
This suppresses a diagram by a factor of $n^3$, leading to graphs
that scale as $\mathcal{O}(1)$ in the large-$n$ limit.
This process is also shown in figure~\ref{fig:2pt_example_single}.
In general it is clear that this procedure generates the melonic-type
graphs shown in figure~\ref{fig:sd1}.

\begin{figure}[t]
	\centering
	\includegraphics[width=0.9\linewidth]{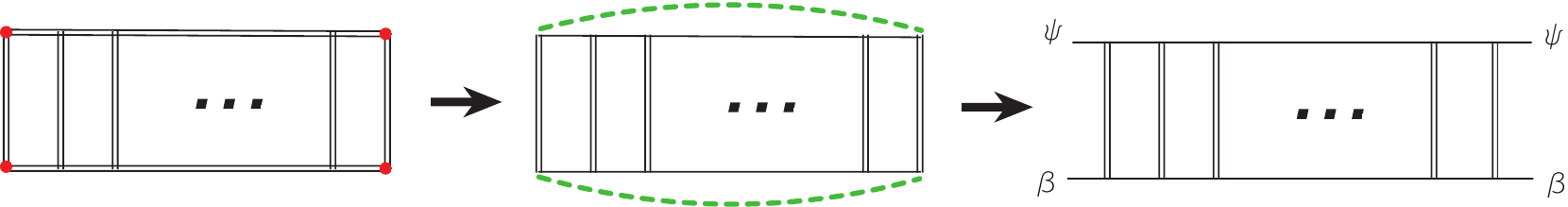}
	\caption{An example showing how to generate dominant contributions of the quark/squark 4-point functions. We start by inserting four vertices, shown in red, into a mesonic vacuum graph. We then unzip horizontally the two pairs of insertions. Finally we cut  the green dashed lines, which come from unzipping the top and the bottom double lines in the previous step, to get the dominant ladder diagrams contributing to the $\<\psi_a(t_1)\b_b(t_2)\psi_a(t_3)\b_b(t_4)\>$ 4-point correlator. }
	\label{fig:4ptgen}
\end{figure}

We can compute leading 4-point function diagrams in a similar manner, but with two quark lines cut. Since the construction is essentially the same we do not elaborate all the details but just show in figure~\ref{fig:4ptgen} one example that generates the dominant ladder diagrams contributing to the 4-point function of some single-line quark fields.

\end{document}